\documentclass[twocolumn]{aastex701}
\usepackage{nicefrac}
\usepackage{microtype}
\usepackage{booktabs}
\usepackage{rotating}
\usepackage{xcolor}
\usepackage[space]{grffile}

\newcommand{\kms}{${\rm km\,s}^{-1}$}

\newcommand{\HE}{HE\,0439$-$5254}

\received{?}
\revised{?}
\accepted{?}
\shorttitle{MUSEQuBES: CGM of an isolated, star-forming dwarf galaxy at $z=0.57$}
\shortauthors{Johnson et al.}

\begin{document}

\title{MUSEQuBES: Physical conditions, origins, and multielement abundances  of the circumgalactic medium of an isolated, star-forming dwarf galaxy at $z=0.57$}

\author[orcid=0000-0001-9487-8583]{Sean D. Johnson}
\affiliation{Department of Astronomy, University of Michigan, 1085 S. University, Ann Arbor, MI 48109, USA}
\email[show]{seanjoh@umich.edu}  
\correspondingauthor{Sean D. Johnson}
\author[orcid=0000-0002-9141-9792]{Nishant Mishra}
\affiliation{Department of Astronomy, University of Michigan, 1085 S. University, Ann Arbor, MI 48109, USA}
\email{mishran@umich.edu}  

\author[orcid=0000-0003-3938-8762]{Sowgat Muzahid}
\affiliation{Inter-University Centre for Astronomy \& Astrophysics, Post Bag 04, Pune 411007, India}
\email{sowgat@iucaa.in}

\author[orcid=0000-0002-8459-5413]{Gwen C. Rudie}
\affiliation{The Observatories of the Carnegie Institution for Science, 813 Santa Barbara Street, Pasadena, CA 91101, USA}
\email{gwen@carnegiescience.edu}

\author[orcid=0000-0001-7869-2551]{Fakhri S. Zahedy}
\altaffiliation{Deceased}
\affiliation{Department of Physics, University of North Texas, Denton, TX 76201, USA}
\email{Fakhri.Zahedy@unt.edu}

\author[orcid=0000-0002-2941-646X]{Zhijie Qu}
\affiliation{Department of Astronomy, Tsinghua University, Beijing, People's Republic of China}
\affiliation{Department of Astronomy \& Astrophysics, The University of Chicago, 5640 S. Ellis Ave., Chicago, IL 60637, USA}
\email{quzhijie@tsinghua.edu.cn}

\author[orcid=0000-0002-4900-6628]{Claude-Andr\'e Faucher-Gigu\`ere}
\affiliation{CIERA and Department of Physics and Astronomy, Northwestern University, Evanston, IL 60201, USA}
\email{cgiguere@northwestern.edu}

\author[orcid=0000-0002-7541-9565]{Jonathan Stern}
\affiliation{School of Physics and Astronomy, Tel Aviv University, Tel Aviv 69978, Israel}
\email{sternjon@tauex.tau.ac.il}

\author[0000-0002-0311-2812]{Jennifer~I-Hsiu Li}
\affiliation{Center for AstroPhysical Surveys, National Center for Supercomputing Applications, University of Illinois Urbana-Champaign, Urbana, IL, 61801, USA}
\affiliation{Michigan Institute for Data Science, University of Michigan, Ann Arbor, MI, 48109, USA}
\affiliation{Department of Astronomy, University of Michigan, Ann Arbor, MI, 48109, USA}
\email{jli184@illinois.edu}

\author[orcid=0000-0001-6846-9399]{Elise Fuller}
\affiliation{Astrophysical \& Planetary Sciences, University of Colorado, Boulder, 2000 Colorado Ave, Boulder, CO 80309, USA}
\affiliation{Department of Astronomy, University of Michigan, Ann Arbor, MI, 48109, USA}
\email{elise.kesler@colorado.edu}

\author[orcid=0000-0001-5804-1428]{Sebastiano Cantalupo}
\affiliation{Department of Physics, University of Milan Bicocca, Piazza della Scienza 3, I-20126 Milano, Italy}
\email{sebastiano.cantalupo@unimib.it}

\author[orcid=0000-0001-8813-4182]{Hsiao-Wen Chen}
\affiliation{Department of Astronomy and Astrophysics, The University of Chicago, Chicago, IL 60637, USA}
\affiliation{Kavli Institute for Cosmological Physics, The University of Chicago, Chicago, IL 60637, USA}
\email{hwchen@uchicago.edu}

\author[orcid=0009-0000-1797-4950]{Ahmad Kadri}
\affiliation{Department of Astronomy, University of Michigan, Ann Arbor, MI, 48109, USA}
\email{amdkadri@umich.edu}

\author[orcid=0000-0003-4427-4831]{Suyash Kumar}
\affiliation{Department of Astronomy and Astrophysics, The University of Chicago, Chicago, IL 60637, USA}
\email{suyashk@uchicago.edu}

\author[0000-0002-2662-9363]{Zhuoqi (Will) Liu}
\affiliation{Department of Astronomy, University of Michigan, 1085 S. University, Ann Arbor, MI 48109, USA}
\email{zql@umich.edu}

\author[orcid=0000-0002-6313-6808]{Gregory Walth}
\affiliation{IPAC, California Institute of Technology, Mail Code 314-6, 1200 E. California Boulevard, Pasadena, CA 91125, USA}
\email{gwalth@ipac.caltech.edu}

\begin{abstract}
In dwarf galaxy models, outflows expel metal-enriched interstellar medium (ISM) into the circumgalactic medium (CGM) to reproduce their observed low metallicities, but measurements of dwarf CGM properties are scarce. We present a study of the CGM of an isolated dwarf at $z=0.5723$ with a stellar mass of $\approx5\times10^7\rm\,M_{\odot}$ and star-formation rate ($\approx0.05\,\rm M_\odot\,yr^{-1}$) and ISM metallicity ($\rm [O/H]\approx-0.9$) consistent with the star-forming main sequence and mass-metallicity relation. A background quasar sightline with archival UV spectra probes the dwarf's CGM at a projected distance of 28 kpc, corresponding to approximately half of the estimated virial radius. The dwarf's CGM is detected in \ion{H}{1}, intermediate metal ions of \ion{C}{3}, \ion{O}{3}, \ion{O}{4}, and \textcolor{black}{\ion{S}{5}}, and kinematically broader, highly-ionized \ion{O}{6}, but is undetected in \ion{N}{4} and \ion{Ne}{8}. Photoionization modeling of the intermediate ions indicates a modest volume-filling factor ($\sim 6\%$ along the sightline or $\sim 2\%$ globally), and a mass of $\sim2\times10^8 {\rm\,M_\odot}$, $\sim4\times$ higher than the dwarf's stellar mass, but $\sim10\times$ less than the highly ionized CGM. The \ion{O}{6} kinematics are comparable to the dwarf's estimated virial velocity, suggesting it is likely associated with cool, photoionized, and volume-filling CGM, with bulk motion or turbulence dominating over thermal pressure. The metallicity inferred for the intermediate ions is $\rm [O/H]=-0.6$, but with low relative abundances of $\rm [C/O]=-0.6$ and \textcolor{black}{$\rm [N/O]<-1.0$}. The [N/O] is below levels expected of the dwarf's ISM, but consistent with core-collapse supernova ejecta, suggesting that supernova-enriched gas escaped the dwarf without mixing significantly with ISM enriched in nitrogen from evolved, low-mass stars.
\end{abstract}



\section{Introduction}
\label{section:introduction}

Low-mass dwarf galaxies exhibit lower stellar-to-halo mass ratios \citep[e.g.,][]{McGaugh:2010, Read:2017} and lower heavy element abundances \citep[e.g.,][]{Lee:2006, Kirby:2013, Berg:2019} than more massive galaxies such as the Milky Way \citep[for reviews, see][]{Bullock:2017, Hunter:2024}. Models of galaxy evolution reproduce these trends by incorporating stellar feedback mechanisms, typically dominated by core-collapse supernovae explosions, that are effective at expelling or heating gas in the shallow gravitational potential wells of dwarf galaxies to slow star formation while also removing heavy elements from the interstellar medium  \citep[ISM; e.g.,][]{Ma:2016, Angles-Alcazar:2017a, Muratov:2017, Emerick:2019, Mina:2021, Pandya:2021, Steinwandel:2024}. Simulations that incorporate this feedback predict that the circumgalactic and intergalactic medium (CGM/IGM) around dwarf galaxies will be a dominant repository for baryons and metals \citep[e.g.,][]{Shen:2014, Vogelsberger:2014, Schaye:2015, Wang:2015, Muratov:2017, Nelson:2019, Baumschlager:2025, Piacitelli:2025}. However, the predicted physical conditions, masses, and heavy element abundances of these reservoirs vary significantly depending on adopted feedback prescriptions \citep[e.g.,][]{Davies:2019, Crain:2023, Faucher-Giguere:2023}. Consequently, observations of the CGM/IGM around dwarf galaxies represent a key means of testing our understanding of galaxy evolution.

In the local Universe, multiphase outflows driving ISM from  galaxies into the nearby CGM can be detected in emission \citep[for a review, see][]{Rupke:2018}, but such observations are not currently feasible around dwarf galaxies at even modest distances. To overcome this obstacle, studies of the diffuse CGM/IGM often rely on background quasar absorption spectroscopy, which is sensitive to gas orders-of-magnitude lower in density than is possible in emission \citep[for reviews, see][]{Tumlinson:2017, Chen:2026}. Absorption observations of the CGM of dwarf galaxies often operate at low redshift ($z<0.05$) due to the difficulty in detecting low-luminosity galaxies at larger distances \citep[e.g.,][]{Bordoloi:2014, Liang:2014, Burchett:2016, Zheng:2020, Qu:2022a, Zheng:2024}. These studies found low covering fractions in low and intermediate ionization state metal lines (e.g, \ion{Si}{2}, \ion{S}{3}, \ion{S}{4}, \ion{C}{4}) in the CGM/IGM around dwarf galaxies, suggesting that their metal and baryon budgets are dominated by gas expelled beyond the virial radius or accounted for by more highly ionized phases. At such low redshifts, however, the limited sensitivity at far-UV wavelengths of instruments on the {\it Hubble Space Telescope} ({\it HST}) prevent observations of the more highly ionized gas traced by the \ion{O}{6} doublet, which is the most common metal transition at low redshift \citep[e.g.,][]{Danforth:2016}. Furthermore, this limited wavelength coverage and the damping wing of the \ion{H}{1} Ly$\alpha$ absorption line arising from the Milky Way ISM prevent measurements of \ion{H}{1} column densities in the CGM/IGM around low redshift dwarfs.

Deep surveys with multi-object spectrographs on large telescopes surveyed faint galaxies in the foreground of UV-bright quasars with archival UV spectra from the Cosmic Origins Spectrograph \citep[COS;][]{Green:2012}, enabling the first statistical studies of the CGM/IGM around dwarf galaxies at modest redshifts \citep[][]{Johnson:2017, Tchernyshyov:2022a}. These studies showed that highly ionized metal lines such as \ion{O}{6} are common in the CGM around dwarf galaxies, though with typical detected column densities that are $\approx0.5$ dex lower than those found around more massive, star-forming galaxies \citep[][]{Chen:2009, Tumlinson:2011, Stocke:2013, Tchernyshyov:2022a, Ho:2025}. Subsequent surveys with state-of-the-art integral field spectrographs (IFS) such as the Cosmic Ultraviolet Baryon Survey \citep[CUBS;][]{Chen:2020} and MUSE Quasar Blind Emitter Survey \citep[MUSEQuBES; ][]{Dutta:2024} pushed galaxy surveys in the fields of UV-bright quasars 2$-$3 magnitudes fainter, dramatically expanding dwarf galaxy samples available for CGM/IGM studies. These new IFS surveys demonstrate that the \ion{O}{6}-bearing component of the CGM (projected distances $d$ less than the estimated dark matter halo virial radius, $R_h$) and nearby IGM ($d/R_{\rm h}=1-2$) are dominant metal reservoirs around low-mass galaxies and exhibit surprisingly quiescent kinematics \citep[e.g.,][]{Mishra:2024, Dutta:2025a, Dutta:2025}.

Recently, high-quality UV absorption spectra at intermediate redshifts of $z=0.4-0.8$ have enabled new insights into the enrichment history of the CGM/IGM through observations of key elements with different nucleosynthetic origins such as  core-collapse supernova (e.g., oxygen and magnesium) versus evolved low-mass stars (e.g,. carbon and nitrogen), and Type Ia supernova \citep[e.g., iron;][]{Lehner:2016, Zahedy:2017a, Zahedy:2021, Cooper:2021, Kumar:2024}. However, studies of the physical conditions and abundances of the CGM around dwarf galaxies are severely limited. Here, we present the first study of the CGM of an isolated, star-forming dwarf galaxy at intermediate redshift, where the \ion{H}{1} Lyman series and a rich multi-element, multi-ion suite of absorption features shift into the observable UV spectral range. The Letter proceeds as follows: In Section \ref{section:observations}, we describe the quasar absorption spectroscopy and follow-up galaxy spectroscopy. In Section \ref{section:results}, we describe the resulting inferences into the properties of the dwarf galaxy and its absorbing CGM. In Section \ref{section:discussion}, we discuss the implications of these results. Finally, in Section \ref{section:conclusions}, we summarize and make concluding remarks.

Throughout, we adopt a flat $\Lambda$ cosmology with $\Omega_{\rm m}=0.3$, $\Omega_{\Lambda} = 0.7$, and a Hubble constant of $H_0= 70\ {\rm km\,s^{-1}\,Mpc^{-1}}$. All magnitudes are in the AB system \citep[][]{Oke:1983}, wavelengths are in vacuum, and projected distances are proper. When characterizing luminosities, we compare to the characteristic knee in the galaxy luminosity function, $L_*$, estimated to correspond to an absolute rest-frame $r$-band magnitude of $M_r=-21.5$ by \cite{Loveday:2012}. When estimating stellar masses and star formation rates (SFRs), we adopt a \cite{Kroupa:2001} initial mass function (IMF).

\begin{figure*}
\includegraphics[width=\textwidth]{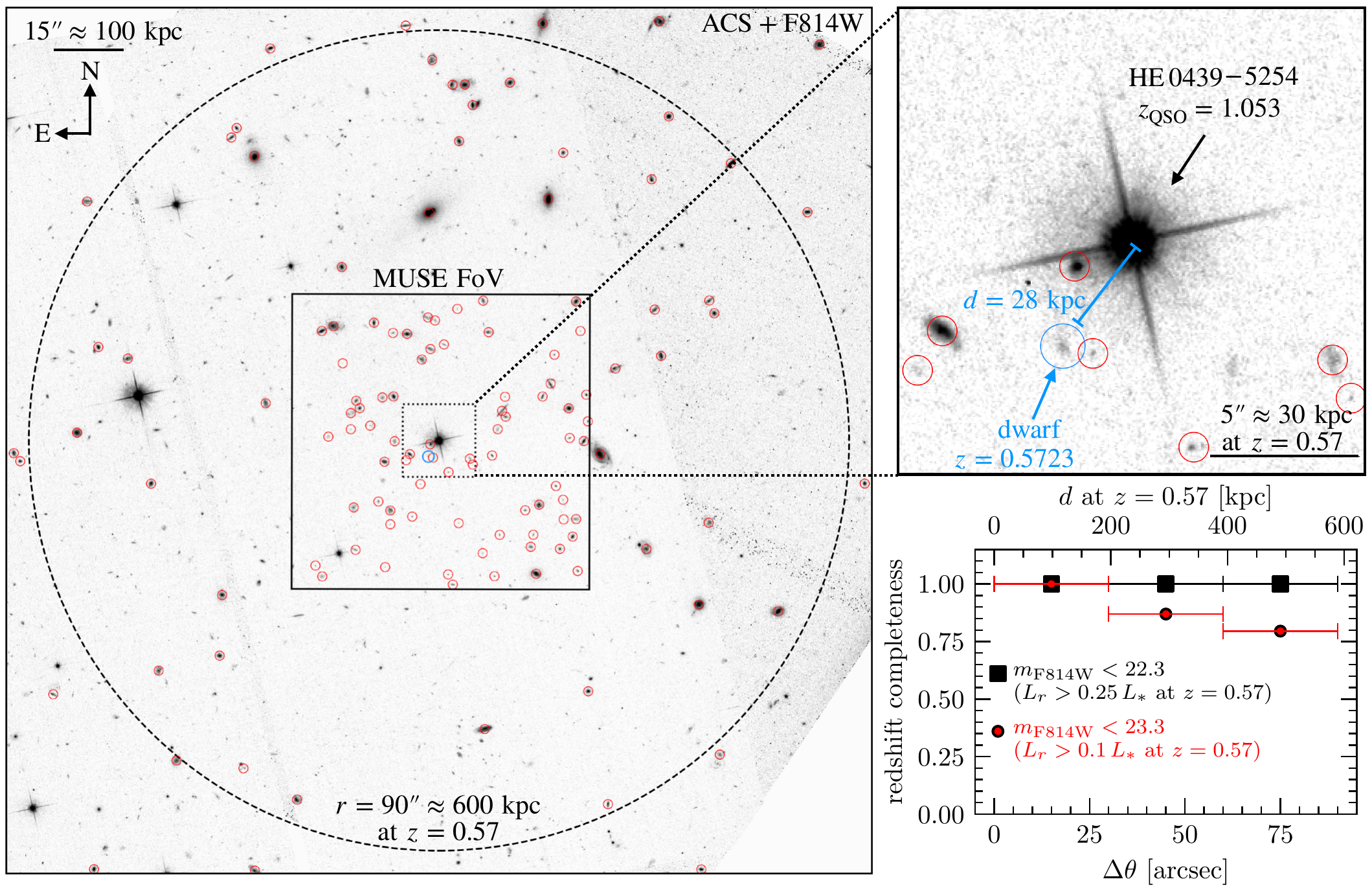}
\caption{Summary of the spectroscopic redshift survey in the field of \HE.  The {\it left} panel displays an archival {\it HST} ACS$+$F814W image of the field, centered on the quasar. The dwarf galaxy at $z=0.5723$ and at a projected distance of $28$ kpc from the quasar sightline is marked by a blue circle while galaxies at other redshifts are marked by smaller red circles. The solid black square shows the MUSE FoV. For scale, the panel also includes a black dashed circle with a radius of $r=90$'', corresponding to $d\approx600$ kpc at $z=0.57$. The top right panel displays a zoom-in on the {\it HST} image to better visualize the dwarf galaxy. The {\it bottom right} panel displays the redshift completeness of the available spectroscopic survey as a function of projected distance from the quasar sightline for galaxies with $m_{\rm F814W}<22.3$ (black squares) and $m_{\rm F814W}<23.3$ (red circles) with angular units on the bottom $x$-axis and the corresponding projected distance at $z=0.57$ on the top $x$-axis. The two magnitude ranges of  $m_{\rm F814W}<22.3$ and $<23.3$ correspond to the expected apparent magnitudes of galaxies brighter than $0.25\, L_*$ and $0.1\, L_*$, respectively. The high spectroscopic completeness levels achieved near the quasar sightline ensure that the dwarf galaxy at $z=0.5723$ is not a satellite of a luminous galaxy.}
\label{figure:survey}
\end{figure*}

\section{Observations and measurements}
\label{section:observations}

\subsection{Galaxy survey observations, data reduction and redshift measurements}

\HE\ is a UV-bright quasar at $z_{\rm QSO}=1.053$ \citep[][]{Wisotzki:2000} targeted by the COS Guaranteed Observing Time (GTO) survey \citep[][]{Stocke:2013}
because it, along with two other quasars, probes the CGM of a local galaxy \citep[][]{Keeney:2013}. To take full advantage of the large redshift path length offered by the $z\approx 1$ quasar and study the CGM/IGM around more distant galaxies, we targeted the field for deep IFS observations and subsequent wide-field follow-up as part of MUSEQuBES \citep[][]{Dutta:2024}.  We observed the field of \HE\ with the Multi Unit Spectroscopic Explorer \citep[MUSE;][]{Bacon:2010}, a wide-field IFS on the 8.4-m Yepun Very Large Telescope (VLT UT4) as part of the MUSE GTO collaboration (PI: Schaye, PID: 094.A-0131). The MUSE observations, conducted in wide-field mode, cover a $1'\times 1'$ FoV centered on the quasar, sampled with $0.2''\times0.2''$ spaxels. The spectra cover a contiguous wavelength range of 4750 \AA\ to $9350$ \AA\ at a resolution of $R\approx 3000$. The observations of \HE\ include a total of 2.5 hours of scientific exposure conducted under full-width-at-half maximum seeing conditions of $\rm FWHM=0.7''$. The MUSE observing strategy and data reduction are detailed in \cite{Muzahid:2021} and \cite{Dutta:2024}.

The field of \HE\ has deep, high angular resolution imaging acquired with the Advanced Camera for Surveys (ACS) aboard {\it HST} as well as deep multi-band {\it griz} ground-based images obtained with the Dark Energy Camera (DECam) on the 4-m Blanco telescope as part of the Dark Energy Survey \citep[DES;][]{Abbott:2021}. Two independent {\it HST} programs observed the field with ACS using the F814W filter (PIs Mulchaey, Lehner; PIDs: 13024, 14269) in Cycles 20 and 23 for a total of 3576 seconds. We retrieved the ACS$+$F814W images from the Mikulski Archive for Space Telescopes (MAST) and combined them using \texttt{SWarp} \citep[][]{Bertin:2002}. The  5$\sigma$ detection limit of the {\it HST} image is $m_{\rm F814W}\approx 27$ for resolved sources with radii of $\approx 0.25''$. The {\it HST} image coverage extends to a minimum of $90''$ from the quasar position in all directions.

To conduct a galaxy redshift survey with the MUSE observations, we first identified sources and measured their photometry in the ACS$+$F814W image with \texttt{Source Extractor} \citep{Bertin:1996} and a white-light image formed from the MUSE datacube. Next, we used \texttt{MPDAF} \cite[][]{Bacon:2016, Piqueras:2017} to extract 1D spectra at the position of each continuum source using circular apertures with radii of between $0.4''$ and $1.6''$, depending on the observed size of the source. We measured a redshift for each source by fitting their MUSE spectra with linear combinations of the first four galaxy eigenspectra from \cite{Bolton:2012} and classified redshifts as secure based on the presence of at least two detected spectral features or as ambiguous if based on a single detected feature.

While deep, the MUSE redshift survey is somewhat narrow and does not cover the full angular scales needed to characterize the group environment of galaxies identified in the MUSE FoV. To conduct a wider-field survey, we targeted galaxies detected in the {\it HST} or DES images within $10'$ of the quasar sightline for spectroscopic follow-up with the LDSS3 and IMACS \citep[][]{Dressler:2011} multi-object spectrographs on the 6.5-m Magellan Telescopes. The wider-field survey targeted galaxies with $i$-band magnitude brighter than 23.5. We observed six multi-slit LDSS3 masks using the VPH-all grism and eight IMACS masks using the 200l grism. We  reduced the data as described in \cite{Cooper:2021} and measured redshifts using the same methods as with MUSE. Comparison of redshifts for galaxies measured with both MUSE and Magellan indicate typical redshift uncertainties of $\approx 50$ \kms\ for LDSS3 and IMACS. The galaxy redshift survey results are summarized in Figure \ref{figure:survey}, which displays the ACS$+$F814W image and the redshift survey completeness levels.

The MUSE survey in the field of \HE\ revealed a faint emission-line source detected in [\ion{O}{2}], H$\beta$, and [\ion{O}{3}] at a redshift of $z=0.5723$ at a right ascension of $04{:}40{:}12.30$ and declination of ${-}52{:}48{:}21.1$ (J2000), $4.3''$ southeast of the quasar sightline, corresponding to a projected distance of $d=28$ kpc.  In the {\it HST} image, the dwarf galaxy is resolved and exhibits a FWHM of $0.4''$ and apparent magnitude of $m_{\rm F814W}=25.4 \pm 0.2$, where the uncertainty includes systematics associated with the choice of aperture and background estimation. At $z=0.57$, this apparent magnitude corresponds to a luminosity of $L\approx 0.01 L_*$. We extracted the dwarf galaxy spectrum from the MUSE datacube using a circular aperture with a radius of $r=0.8''$, which maximizes the signal-to-noise ratio in the detected emission lines.

While the dwarf galaxy is detected in both emission lines and continuum in the MUSE data, broad wings on the MUSE point spread function result in nonnegligible contamination of the dwarf galaxy's continuum by light from the quasar. To estimate and subtract the quasar contribution, we identified eighteen source-free locations at the same angular separation from the quasar as the dwarf galaxy. We then extracted the spectra from these regions using circular apertures with an $r=0.8$'' radius, computed the mean source-free spectrum, and subtracted it from the dwarf galaxy extraction. Finally, we performed aperture corrections as a function of wavelength based on the curve of growth measured for the quasar and corrected for Milky Way extinction following \cite{Schlafly:2011}.

To characterize the environment of the dwarf galaxy, we identified all galaxies in our spectroscopic   survey with a line-of-sight velocity of $|\Delta v|<500$ \kms\ from the dwarf's systemic redshift. The nearest galaxies to the dwarf in our survey are at angular separations of $\gtrsim 5.8$ arcmin from the quasar sightline, corresponding to projected distances of $>2$ Mpc. Moreover, the high completeness levels of the survey rule out the presence of galaxies brighter than $m_{\rm F814W}=22.3$ (23.3) within $90''$ ($30''$), which corresponds to galaxies more luminous than $L=0.25 L_*$ ($0.1 L_*$) at projected distances of $d<600$ ($<200$) kpc at $z=0.57$, indicating that the dwarf galaxy near the sightline of \HE\, is isolated.

\begin{figure*}
\includegraphics[width=\textwidth]{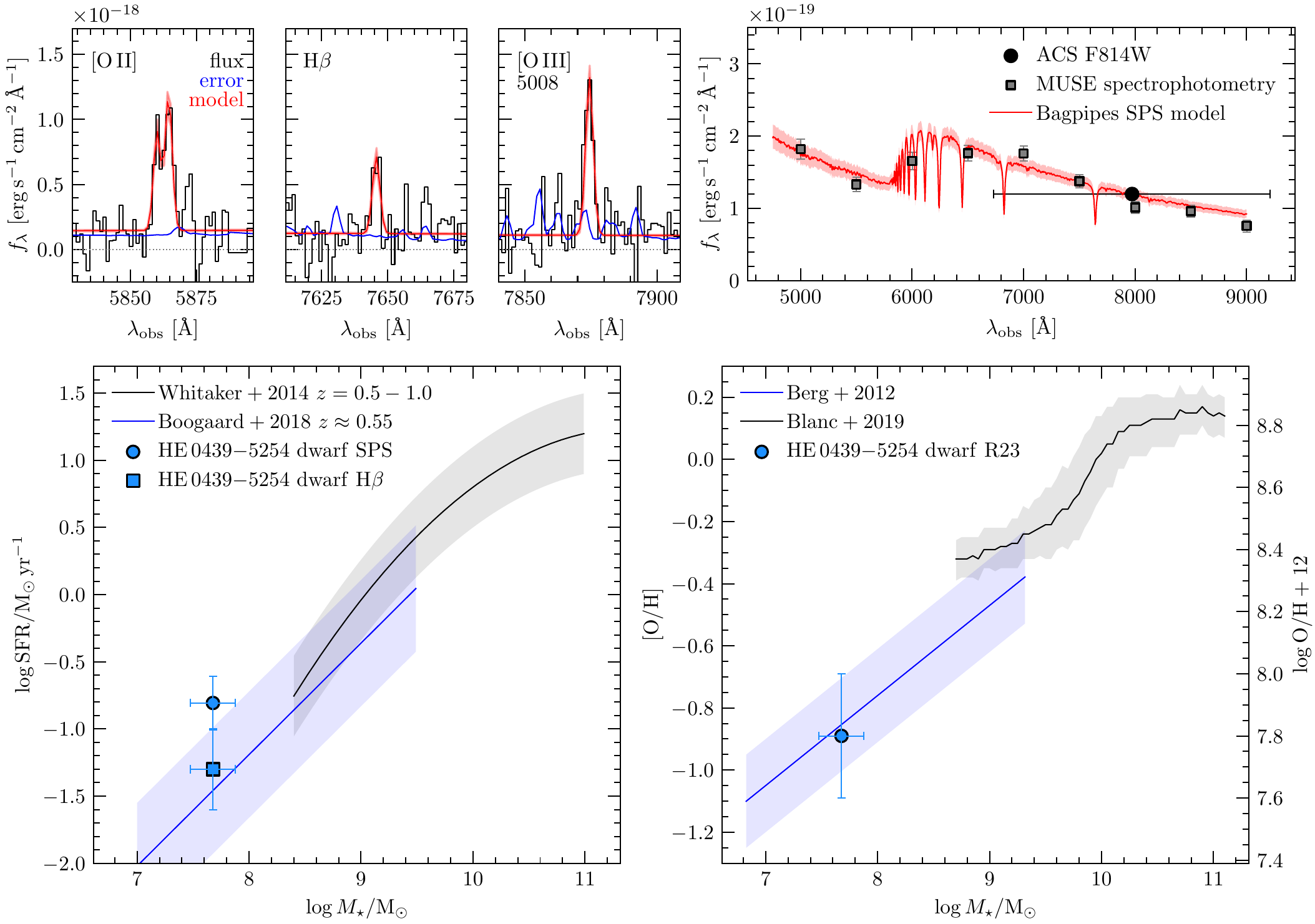}
\caption{Summary of the properties of the dwarf galaxy in the field of \HE. The {\it top} panels display observed spectroscopic and photometric measurements of the dwarf with zoom-ins on the [\ion{O}{2}], H$\beta$, and [\ion{O}{3}] emission lines from MUSE in the {\it left} three panels and broader spectral energy distribution (SED) from both MUSE and the {\it HST} ACS$+$F814W image on the {\it right}. The emission-line spectral panels display the MUSE spectrum without rebinning while the SED plot shows the MUSE spectrum rebinned to 500 \AA\ intervals. The red line in the emission-line panels on the left marks the best-fitting Gaussian emission line and local continuum models. The red line in the SED panel on the right shows the best-fitting \texttt{Bagpipes} SPS model. In both cases, the faded red bands mark the 68\% credible intervals calculated from model fit posteriors. The {\it bottom left} panel displays the inferred star-formation rate and stellar mass of the dwarf galaxy (blue points) in comparison to the mean trend among star-forming galaxies at similar redshift from \cite{Boogaard:2018} and  \cite{Whitaker:2014}. In both cases, the shaded bands mark intrinsic scatter in the relations. The {\it bottom right} panel displays the inferred ISM metallicity and stellar mass of the dwarf galaxy in comparison to the low-redshift mass-metallicity relations from \cite{Berg:2012} and \cite{Blanc:2019}. The star formation rate and metallicity of the dwarf galaxy are consistent with those expected from the star-forming main sequence and mass-metallicity relation.}
\label{figure:spectrum}
\end{figure*}

\subsection{Quasar absorption spectroscopy}
The COS GTO survey observed \HE\  for a total of 8403 seconds with the G130M grating and 8946 seconds with the G160M grating, providing contiguous observed wavelength coverage from $\lambda=1140$ to $1780$ \AA\  (PI: Green, PID: 11520). We retrieved these archival COS spectra from MAST as 1D spectra from individual exposures reduced with \texttt{CalCOS}  version 3.4.3. We combined the individual exposures into a single coadded spectrum after correcting for systematic uncertainty in the COS wavelength calibration by calculating offsets between exposures using all available, well-detected Milky Way and extragalactic absorption features as described in \cite{Johnson:2014} and \cite{Chen:2020}. Finally, we computed the flux uncertainty in each pixel based on Poisson photon counting statistics \citep[e.g.,][]{Gehrels:1986} and performed continuum normalization via low-order spline fitting to feature-free regions of the spectrum. The coadded COS spectrum exhibits a median signal-to-noise ratio of $S/N=16$ per $18$ \kms\ resolution element.

\section{Results}
\label{section:results}

\subsection{Dwarf galaxy properties}
To infer emission-line-based properties of the dwarf galaxy, including a star-formation rate and ISM metallicity, we fit the [\ion{O}{2}] $\lambda\lambda 3727, 3729$ doublet, H$\beta$, and the [\ion{O}{3}] $\lambda\lambda 4960, 5008$ doublet with local linear continua and single-component Gaussian profiles with shared redshift and line-width, accounting for the MUSE line spread function measured by \cite{Bacon:2017}. We used the \texttt{LMFIT} package \citep[][]{Newville:2016} to find the best-fit model and estimated uncertainties with a Markov Chain Monte Carlo (MCMC) approach with \texttt{emcee} \citep[][]{Foreman-Mackey:2013}. The resulting measured fluxes in the [\ion{O}{2}] doublet, H$\beta$, and [\ion{O}{3}] $\lambda 5008$ lines are $7.5\pm 0.5$, $2.5\pm0.3$, and $5.2\pm0.5$, respectively, in units of $\rm 10^{-18}\,erg\,s^{-1}\,cm^{-2}$. The low, marginally resolved, 1D velocity dispersion of $38\pm 6$ \kms\ measured for the dwarf galaxy's emission lines is consistent with expectations for low-mass galaxies \citep[][]{Forbes:2011}. The emission-line spectra and model fits are displayed in the top left panels of Figure \ref{figure:spectrum}.

To estimate properties of the dwarf galaxy from these emission line measurements, we converted the H$\beta$ line flux to luminosity, $L({\rm H}\beta) = 3.3\times10^{39}\ {\rm erg\,s^{-1}}$. Adopting the $L({\rm H\beta})-\rm SFR$ relation from \cite{Boogaard:2018} for MUSE-detected galaxies of similar mass and redshift, the inferred star-formation rate of the dwarf is $\log {\rm SFR}/{\rm M_\odot\,yr^{-1}}\approx -1.3 \pm 0.3$. While Balmer absorption is negligible, we note that the H$\beta$-based SFR estimate is not corrected for dust extinction and therefore represents a lower limit. Furthermore, we measured the O32 ionization diagnostic line ratio, $\log{\rm O23} \equiv \log (\frac{\rm [O\,III]\ \lambda 5008}{\rm [O\,II] \lambda\lambda 3727,3729})=-0.16 \pm 0.05$, and the R23 metallicity diagnostic line ratio, $\log{\rm R23} \equiv \log (\frac{\rm [O\,III]\ \lambda\lambda 4960, 5008 + [O\,II]\ \lambda\lambda 3727,3729}{\rm H\beta})=0.75 \pm 0.06$. We recalibrated the R23 ratio$-$metallicity relation \citep[for a review, see][]{Maiolino:2019} to account for dwarf galaxies with moderate ionization states similar to that observed in this work (see Appendix \ref{appendix:R23} and Figure \ref{figure:R23}). The ISM metallicity of the dwarf galaxy inferred from its R23 ratio is $\log \rm O/H+12 = 7.8 \pm 0.2$ or $\rm [O/H] =-0.9\pm 0.2$ when expressed relative to the solar oxygen abundance from \cite{Asplund:2009}.

\begin{figure*}
\centering
\includegraphics[width=1\textwidth]{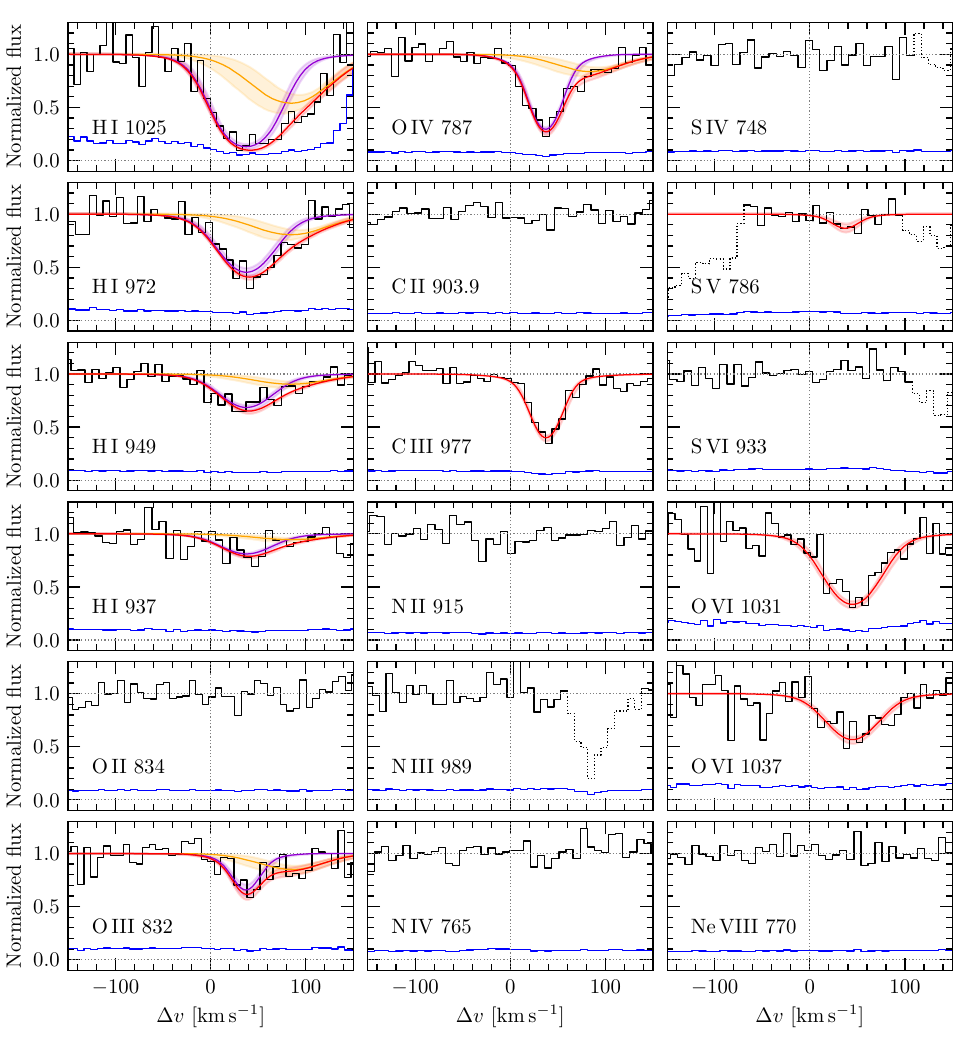}
\caption{Absorption associated with the dwarf galaxy CGM as a function of line-of-sight velocity, $\Delta v$, relative to the dwarf galaxy's systemic redshift, $z=0.5723$. Each panel displays the continuum-normalized spectrum in black histogram, error array in blue, and best-fit Voigt model described in Section \ref{section:absorption} in red line. The COS spectra are binned by a factor of three for visualization purposes. The best fit model for the main component and wing are shown in violet and orange line, respectively. The faded band around the best fit model and each component represent the 68\% credible interval of the Voigt profile model fit posterior. The spectrum around the \ion{N}{4} $\lambda 765$ line is contaminated by \ion{S}{4} $\lambda 744$ at $z=0.6150$. However, the contamination is well constrained by the stronger \ion{S}{4} $\lambda748$ line. Similarly, the \ion{H}{1} Ly$\beta$ absorption associated with the dwarf CGM is contaminated by \ion{H}{1} Ly$\alpha$ from a strong system at $z=0.327$ that is constrained by higher order Lyman series lines. \textcolor{black}{To infer absorption levels from the dwarf CGM in the presence of this contamination, we simultaneously modeled the contaminating features and dwarf CGM as described in Section \ref{section:absorption} and Appendix \ref{appendix:contamination}. For visualization purposes only, we then divided the continuum-normalized quasar spectrum by the model for the contaminating systems and display the results in the H\,I $\lambda$1025 and N\,IV $\lambda$765 panels.} Spectral regions contaminated by strong intervening features that cannot be modeled are shown in dotted line.}
\label{figure:absorption}
\end{figure*}

To characterize the stellar mass and stellar population of the dwarf galaxy, we simultaneously modeled its MUSE spectrum and ACS$+$F814W photometry with the \texttt{Bagpipes} \citep[][]{Carnall:2018} Stellar Population Synthesis (SPS) package. We adopted a double power-law star formation history parameterization with priors described in \cite{Carnall:2019} and a \cite{Calzetti:2000} dust extinction law. The inferred stellar mass of the dwarf galaxy from the SPS modeling is $\log M_\star/{\rm M_\odot} = 7.7\pm 0.2$. Adopting the stellar-to-halo mass relation from \cite{Kravtsov:2018} and virial radius definition from \cite{Bryan:1998}, the estimated halo mass is $\log M_{\rm h}/M_\odot \approx 10.4$ and virial radius is $R_{\rm h}\approx 60$ kpc. The inferred star-formation rate averaged over the last 100 million years is $\log {\rm SFR}/{\rm M_\odot\,yr^{-1}} = -0.8 \pm 0.2$,  mass-weighted mean stellar age is $\log {\rm age}/{\rm yr}\approx 8.4 \pm 0.2$, and dust extinction is $A_V = 0.1^{+0.15}_{-0.08}$. The ACS$+$F814W photometry and MUSE spectrophotometry are compared to the SPS model in the top right panel of Figure \ref{figure:spectrum} after rebinning the MUSE spectrum to 500 \AA\ bins for visualization purposes. The star-formation rate inferred from H$\beta$ is 0.5 dex (a factor of $\approx 3$) lower than the SED-based estimate, but these estimates are fairly consistent given the large systematic uncertainties and lack of a dust correction for H$\beta$. The optical size  and star-formation rates lead to an estimated star-formation rate surface density of $\log {\Sigma_{\rm SFR}}/ M_\odot\,{\rm yr^{-1}\,kpc^{-2}} \approx -2.4$.

Figure \ref{figure:spectrum} shows that the dwarf galaxy's star-formation rate, ISM metallicity, and stellar mass estimates are consistent with expectations from the observed correlation between stellar mass and star formation rate \citep[often referred to as the ``star-forming main sequence''; e.g.,][]{Whitaker:2014, Boogaard:2018} and the low-redshift mass-metallicity relation \citep[e.g.,][]{Berg:2012, Blanc:2019}.
The dwarf therefore falls within expectations for galaxies of similar mass, indicating that it is a fairly typical star-forming dwarf galaxy at this epoch.

\begin{table*}\
\centering
\caption{Absorption Voigt profile fitting results}
\label{table:absorption}
\begin{tabular}{clccccl}
\hline \hline
component & \multicolumn{1}{c}{ion \textcolor{black}{(transitions)}} & $z$ & \multicolumn{1}{c}{$\Delta v$\tablenotemark{a}}  &  \multicolumn{1}{c}{$b_{\rm turb}$} & $\log T/{\rm K}$ & \multicolumn{1}{c}{$\log N/{\rm cm^{-2}}$} \\
                  &                                       &        &   \multicolumn{1}{c}{[${\rm km\,s^{-1}}$]}  & \multicolumn{1}{c}{[${\rm km\,s^{-1}}$]}  \\

\hline
main &     & $0.57250 \pm 0.00001$ & $38$ & $15 \pm 2$ & $4.6\pm0.2$ &  \\
   & \ion{H}{1}\ \ \ \textcolor{black}{($\lambda$1025, $\lambda$972, $\lambda$949, $\lambda$937)}  & & & & &  $\,\,\,\,\,14.88 \pm 0.05$\\
   & \ion{O}{2}\ \ \textcolor{black}{($\lambda$834)}                                               & & & & &  ${<}13.3$ \\
   & \ion{O}{3} \textcolor{black}{($\lambda$832)}                                                  & & & & &  $\,\,\,\,\,13.92 \pm 0.09$ \\
   & \ion{O}{4} \textcolor{black}{($\lambda$787)}                                                  & & & & &  $\,\,\,\,\,14.46 \pm 0.04$ \\
   & \ion{C}{2}\ \ \textcolor{black}{($\lambda 903$b)}                                             & & & & &  ${<}12.9$ \\
   & \ion{C}{3} \textcolor{black}{($\lambda 977$)}                                                 & & & & &  $\,\,\,\,\, 13.39\pm 0.04$ \\
   & \ion{N}{2}\ \ \textcolor{black}{($\lambda 915$)}                                              & & & & &  ${<}13.0$ \\
   & \ion{N}{3} \textcolor{black}{($\lambda$989)}                                                  & & & & &  ${<}13.5$ \\
   & \ion{N}{4} \textcolor{black}{($\lambda$765)}                                                  & & & & &  ${<}12.7$ \\
   & \textcolor{black}{\ion{S}{2}\ \ ($\lambda 765$)}                                              & & & & &  \textcolor{black}{${<}12.7$} \\
   & \textcolor{black}{\ion{S}{3}\ ($\lambda 724$)}                                              & & & & &  \textcolor{black}{${<}13.4$} \\
   & \textcolor{black}{\ion{S}{4}\ ($\lambda 748$)}                                              & & & & &  \textcolor{black}{${<}12.9$} \\
   & \textcolor{black}{\ion{S}{5}\ \ ($\lambda 786$)}                                              & & & & &  \textcolor{black}{$\,\,\,\,\, 12.35\pm 0.16$} \\
   & \textcolor{black}{\ion{S}{6}\ ($\lambda 933$)}                                              & & & & &  \textcolor{black}{${<}12.8$} \\
wing &   & $0.57275 \pm 0.00004$  & $85$ &  $30 \pm 9$     & $4.4\pm0.3$  &  \\
   & \ion{H}{1}\ \ \ \textcolor{black}{($\lambda$1025, $\lambda$972, $\lambda$949, $\lambda$937)}  & & & & &  $\,\,\,\,\,14.30\pm 0.13$ \\
   & \ion{O}{2}\ \ \textcolor{black}{($\lambda$834)}                                               & & & & &  ${<}13.4$ \\
   & \ion{O}{3} \textcolor{black}{($\lambda$832)}                                                  & & & & &  $\,\,\,\,\,13.61 \pm 0.15$  \\
   & \ion{O}{4} \textcolor{black}{($\lambda$787)}                                                  & & & & &  $\,\,\,\,\,13.71 \pm 0.11$  \\
   & \ion{C}{2}\ \ \textcolor{black}{($\lambda 903$b)}                                             & & & & &  ${<}12.8$ \\
   & \ion{C}{3} \textcolor{black}{($\lambda 977$)}                                                 & & & & &  ${<}12.8$ \\
   & \ion{N}{2}\ \ \textcolor{black}{($\lambda 915$)}                                              & & & & &  ${<}13.0$ \\
   & \ion{N}{4} \textcolor{black}{($\lambda$765)}                                                  & & & & &  ${<}12.6$ \\
   & \textcolor{black}{\ion{S}{2}\ \ ($\lambda 765$)}                                              & & & & &  \textcolor{black}{${<}12.7$} \\
   & \textcolor{black}{\ion{S}{3}\ ($\lambda 724$)}                                                & & & & &  \textcolor{black}{${<}13.3$} \\
   & \textcolor{black}{\ion{S}{4}\ ($\lambda 748$)}                                                & & & & &  \textcolor{black}{${<}12.9$} \\
   & \textcolor{black}{\ion{S}{5}\ \ ($\lambda 786$)}                                              & & & & &  \textcolor{black}{${<}12.2$} \\
   & \textcolor{black}{\ion{S}{6}\ ($\lambda 933$)}                                                & & & & &  \textcolor{black}{${<}12.9$} \\
\hline
component & \multicolumn{1}{c}{ion (transitions)} & $z$ & $\Delta v$ & $b$ & $\log T/{\rm K}$ & \multicolumn{1}{c}{$\log N/{\rm cm^{-2}}$} \\
                  &                                       &        &   \multicolumn{1}{c}{[${\rm km\,s^{-1}}$]}  & \multicolumn{1}{c}{[${\rm km\,s^{-1}}$]}  \\
\hline
high ions & &  $ 0.57253 \pm 0.00002$ & $44$ & $31 \pm 4$ & $-$ \\
                 & \ion{O}{4}\ \ \ \,\,\,\textcolor{black}{($\lambda$787)} &                                         &                   &       &        &  ${<}14.2$ \\
                  & \ion{O}{6}\ \ \ \ \,\textcolor{black}{($\lambda\lambda$1031, 1037)} & & & &  & $\,\,\,\,\,14.32 \pm 0.04$ \\
                 & \ion{Ne}{8} \textcolor{black}{($\lambda$770)} &                                      &                           &  &  & ${<}13.4$  \\
                & \textcolor{black}{\ion{S}{6}\ \ \ \ \,($\lambda 933$)}                                                & & & & &  \textcolor{black}{${<}12.9$} \\

\hline
\multicolumn{7}{l}{\tablenotemark{a}Line-of-sight component velocity relative to the dwarf galaxy at $z=0.5723$}
\end{tabular}
\end{table*}

\subsection{Absorbing circumgalactic medium}
\label{section:absorption}
The archival COS spectra of \HE\ provide an opportunity to study the CGM of the dwarf galaxy at a projected distance of 28 kpc, which corresponds to approximately half of the dwarf galaxy's estimated virial radius. At $z=0.5723$, a suite of absorption features including the \ion{H}{1} Lyman series and ions of carbon, nitrogen, and oxygen shift into the observed spectral window.  The dwarf galaxy's CGM is detected in strong \ion{H}{1} absorption in \ion{H}{1} Ly$\beta, \gamma$, and $\delta$, and metal ions including \ion{O}{3} $\lambda 832$, \ion{O}{4} $\lambda 787$, \ion{C}{3} $\lambda 977$, and the \ion{O}{6}  $\lambda\lambda 1031, 1037$ doublet, \textcolor{black}{as well as a weak detection of \ion{S}{5} $\lambda$786}. The continuum-normalized absorption spectra for each of these transitions are plotted as a function of line-of-sight velocity relative to the dwarf galaxy ($\Delta v$)  in Figure \ref{figure:absorption}. The red side of the \ion{H}{1} Ly$\beta$ associated with the dwarf \textcolor{black}{(at velocities $\Delta v\gtrsim 70$ \kms)} is contaminated by Ly$\alpha$ from a strong \ion{H}{1} system at $z\approx 0.327$ which is well constrained by higher order Lyman series lines \textcolor{black}{(see Appendix \ref{appendix:contamination})}.

The dwarf galaxy's CGM exhibits no detected absorption in \ion{O}{2}, \textcolor{black}{\ion{C}{2}, \ion{N}{2} \ion{N}{3}, \ion{S}{4}, \ion{S}{6}} and \ion{Ne}{8}, which are also plotted in Figure \ref{figure:absorption}. The COS spectra cover the \ion{N}{4} $\lambda765$ line, but this spectral region \textcolor{black}{(at velocities $\Delta v \approx -60$ to $+140$ \kms)} is contaminated by intervening \ion{S}{4} $\lambda744$ absorption at $z=0.6150$. The contamination levels are constrained by the stronger \ion{S}{4} $\lambda748$ line \textcolor{black}{(see Appendix \ref{appendix:contamination})}. To visualize potential \ion{N}{4} absorption from the dwarf galaxy's CGM \textcolor{black}{in Figure \ref{figure:absorption}}, we fit the \ion{S}{4} $\lambda748$ feature at $z=0.6150$ and divided the spectrum by the resulting prediction for the contaminating \ion{S}{4} $\lambda744$ absorption. After removal of the \ion{S}{4} contamination, the CGM of the dwarf galaxy exhibits no detectable \ion{N}{4} absorption (see Figure \ref{figure:absorption}). The dwarf's CGM is also undetected in \ion{N}{3} $\lambda 989$, though with some contamination that cannot be modeled.

While the absorption features associated with the dwarf galaxy's CGM are relatively simple, each consisting of a dominant main component at $z\approx 0.5725$, the \ion{O}{4} profile exhibits a significant non-Gaussian wing, $\approx 40$ \kms\ redward of the stronger component. This wing is also seen in less prominent asymmetries observed in \ion{H}{1}, and possibly \ion{O}{3}. To measure the column densities as well as thermal and non-thermal broadening levels implied by the observed absorption, we performed a joint two-component Voigt profile fit to the \ion{H}{1}, \ion{O}{3} $ \lambda 832$, \ion{O}{4} $\lambda 787$, \ion{C}{3} $\lambda 977$, and \ion{S}{5} $\lambda786$ features with one component for the main absorber and one for the wing. The model parameters for each component include redshift, the \ion{H}{1} and metal ion column densities, temperature ($T$), and turbulent, non-thermal broadening ($b_{\rm turb}$). The temperature and turbulent broadening effectively set the line widths of the model for each absorption feature of a component as $b = \sqrt{b^2_{\rm therm}(T, m)+ b^2_{\rm turb}}$, with $b_{\rm therm}(T, m) = \sqrt{\frac{2 k_{\rm B} T}{m}}$ representing thermal Doppler broadening where $m$ is the mass of the element and $k_B$ is the Boltzmann constant.

The \ion{O}{6} absorption is significantly broader than \ion{O}{3} and \ion{O}{4} features, so we fit the doublet with a distinct, single highly ionized Voigt profile component with no attempt to differentiate thermal and non-thermal broadening. At fixed metallicity, highly ionized \ion{O}{6} absorbers are expected to produce significantly less \ion{H}{1} absorption than components with strong detections of \ion{C}{3}, \ion{O}{3}, and \ion{O}{4}. However, without observations of Ly$\alpha$ associated with the CGM of the dwarf, we cannot rule out the possibility of nonnegligible contributions to the total \ion{H}{1} from the \ion{O}{6}-bearing gas. If the highly ionized component contributes nonnegligible \ion{H}{1} absorption, then metallicity estimates of the intermediate ions would represent lower limits while total column and temperature estimates would represent upper limits.

To account for the contamination of \ion{H}{1} Ly$\beta$ by the strong \ion{H}{1} system at $z=0.327$, \textcolor{black}{we simultaneously fit the dwarf CGM features and contaminating absorption using constraints from Lyman series lines as described in Appendix \ref{appendix:contamination}}.  We note that the contaminating \ion{H}{1} Ly$\alpha$ complicates continuum fitting around the Ly$\beta$ line, which is also saturated. \textcolor{black}{To ensure that continuum uncertainty and saturation of Ly$\beta$ do not bias the inferred CGM \ion{H}{1} properties, we performed an independent fit to the dwarf CGM \ion{H}{1} that excluded Ly$\beta$ and found measured properties consistent within uncertainties for both the main component and wing.} To determine upper limits based on undetected features associated with each component, the Voigt profile model includes absorption from potential \ion{O}{2}, \ion{N}{3}, and \ion{N}{4} for the main component as well as \ion{O}{2}, \ion{C}{3} and \ion{N}{4} for the wing. To account for the contamination of \ion{N}{4} by the intervening \ion{S}{4} $\lambda 744$ absorption at $z=0.6150$, we simultaneously fit a multi-component model for the contamination, \textcolor{black}{as described in Appendix \ref{appendix:contamination}}. \textcolor{black}{Simultaneously modeling the contaminating features ensures that additional uncertainty due to the contamination are included in the H\,I measurements and \ion{N}{4} upper limits}. To place limits on the presence of other ions associated with the highly ionized \ion{O}{6}-bearing component, we included potential absorption from \ion{O}{4} $\lambda 787$, \textcolor{black}{\ion{S}{6} $\lambda$933,} and \ion{Ne}{8} $\lambda 770$  with width and redshift tied to the \ion{O}{6} absorption.

We convolved the full Voigt model including the main component, wing, and highly ionized component  with the wavelength-dependent COS line spread function from lifetime position 1 and performed $\chi^2$ minimization with \texttt{LMFIT} and MCMC posterior exploration with \texttt{emcee}. We chose a prior for column densities that is bounded from below by zero and flat in linear column, $N$. A flat prior in $N$ is uninformative for detections, producing results consistent with those from a flat prior in $\log$, and it prevents convergence issues for nondetections that arise with log priors due to $\log N/{\rm cm^{-2}}$ approaching $-\infty$ as $N$ approaches zero. The best-fit model and contribution from each component are plotted along with the observed absorption in Figure \ref{figure:absorption}, and the resulting absorption line measurements are summarized in Table \ref{table:absorption}. \textcolor{black}{Table \ref{table:absorption} also notes the transitions used for each measurement or upper limit}. For all nondetections, we report 95\% upper limits on the associated columns with temperature and non-thermal broadening set by each component's detected features. For all analysis involving the nondetections, we use the posterior for the column densities marginalized over all other model parameters, negating the need to choose a detection threshold in the likelihood.

\subsection{Ionization modeling}
\label{section:ionization}

\begin{figure*}
\includegraphics[width=\textwidth]{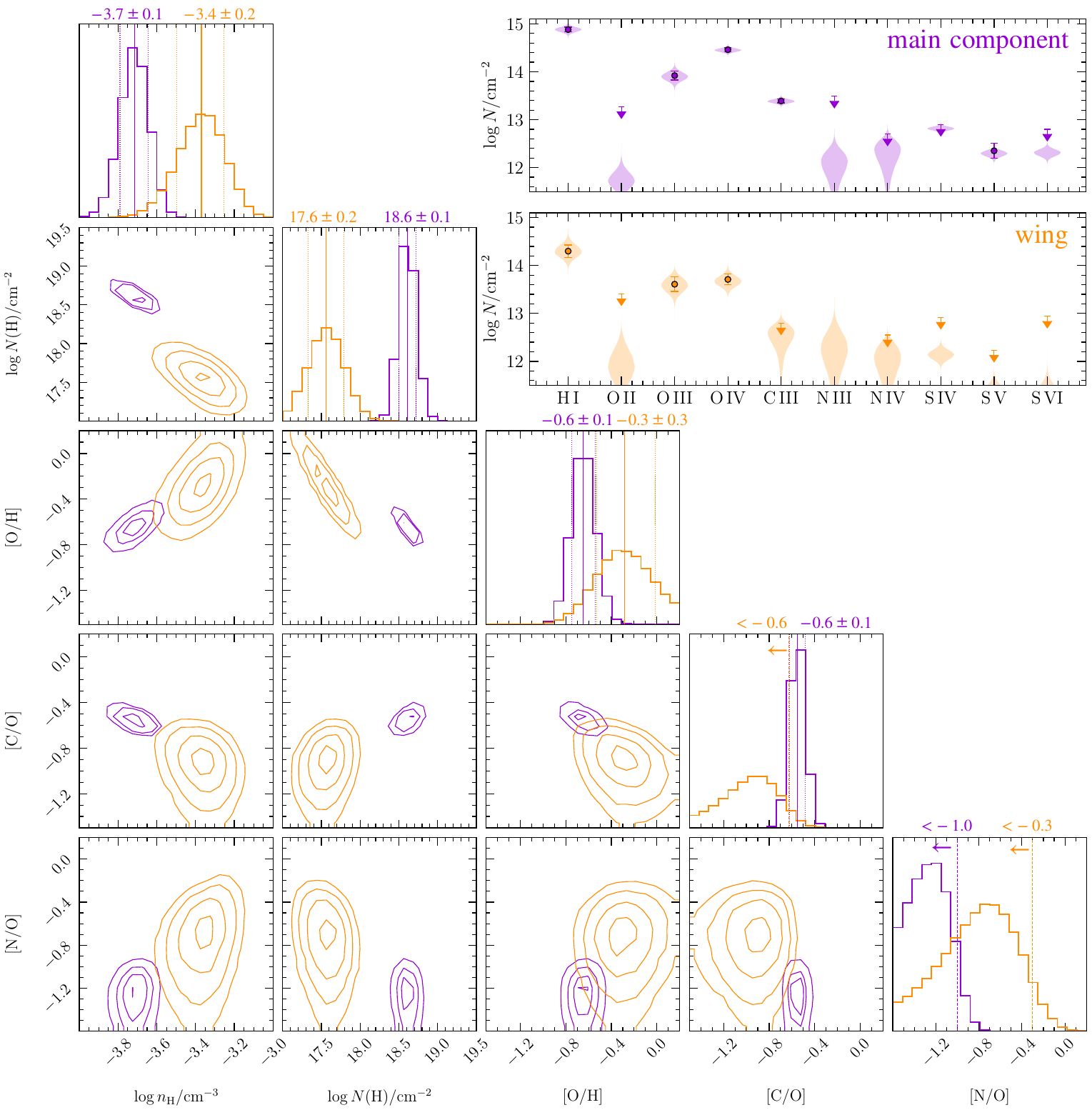}
\caption{Summary of the \texttt{Cloudy} photoionization equilibrium model fits to the main component and wing shown in violet and orange, respectively. The corner plot occupying the bottom left half of the figure visualizes the two dimensional covariances between model parameter pairs and the one-dimensional histograms at the top of each column show the posterior for each parameter marginalized over all others. For well-measured parameters, the median and 68\% credible interval are visualized by solid and dotted vertical line in the one-dimensional histograms, and these are also given in text above each histogram. In the case of upper limits on parameters, a vertical dashed line marks the 95\% upper limit, which is also labeled. The two violin plots in the top right of the figure display the measured column densities as data points with error bars or 95\% upper limits marked with downward arrows for \ion{H}{1}, \ion{O}{2}, \ion{O}{3}, \ion{O}{4}, \ion{N}{3}, and \ion{N}{4} for the main component (top, violet) and wing (bottom, orange) along with the photoionization model column density posteriors in shaded bands. \textcolor{black}{We note that the posteriors for some ions in the violin plot are necessarily cut-off because they extend to minus infinity or more than an order of magnitude below the upper limits. The violin plot does not include \ion{C}{2} or \ion{N}{2} because these nondetections are several orders of magnitude too weak to constrain the ionization models.}}
\label{figure:cloudy}
\end{figure*}

To infer physical conditions and heavy element abundances of the dwarf galaxy's CGM from the absorption measurements described in Section \ref{section:absorption}, we performed ionization modeling using \texttt{Cloudy} version 23.01 \citep[][]{Chatzikos:2023}. For the main component and wing, which are dominated by intermediate ionization states, we assume photoionization equilibrium with the UV background from \cite{Faucher-Giguere:2020} evaluated at $z=0.57$ and produced a grid of simulated clouds with varying hydrogen density ($n_{\rm H}$), total hydrogen column density ($N_{\rm H}$), and metallicity ($\rm [O/H]$) with solar abundance patterns from \cite{Asplund:2009}. The temperature of each photoionization equilibrium model cloud, $T_{\rm PIE}$,  is set by the balance of photoheating and metallicity-dependent radiative cooling. We ran a grid of \texttt{Cloudy} models covering $\log n_{\rm H}/{\rm cm}^{-3}=-5$ to $-1$, $\log N_{\rm H}/{\rm cm}^{-2}=16$ to $21$, and $\rm [O/H] = -2$ to $+0.3$ in steps of $0.1$ dex.  We then interpolated the resulting grid to produce predicted column densities for \ion{H}{1} and all observed metal ions as a function of density, total hydrogen column density, and oxygen metallicity. \textcolor{black}{We note that inclusion of local ionizing radiation from the dwarf's stellar population does not significantly alter the model predictions, even with an escape fraction of $f_{\rm esc}=1$. The local ionization spectrum from young stars in the dwarf could be important for inner CGM at $\lesssim15$ kpc distance from the sites of star formation.}

\textcolor{black}{To enable constraints on CGM chemistry, we included scaling factors to allow for nonsolar [C/O] and [N/O] abundance patterns because a significant fraction of carbon and nitrogen arises from winds driven by evolved, low-mass stars} \textcolor{black}{while oxygen arises primarily from core-collapse supernovae ejecta}. nonsolar carbon and nitrogen abundances do not significantly change PIE ion fractions because carbon and nitrogen are sub-dominant contributors to the cooling function \citep[also see][]{Kumar:2025}.
\textcolor{black}{On the other hand, we adopted Solar [S/O] relative abundances. The majority of sulfur is thought to arise from winds driven by core-collapse supernovae, though with nonnegligible contributions from Type Ia supernovae \citep[e.g.,][]{Kobayashi:2020}, and some galaxies at $z\approx 3$ exhibit sub-solar [S/O] consistent with a lack of Type Ia nucleosynthetic products \citep[][]{Rogers:2024, Rogers:2025}. However, correlations between [S/O] abundances and metallicities observed in low-mass, low-redshift galaxies \citep[e.g.,][]{Thuan:1995, Esteban:2025} are sufficiently weak that departures from solar [S/O] abundances due to varying Type Ia contributions are not expected to be detectable given the large uncertainty in the \ion{S}{5} column density observed in the dwarf's CGM.}

Given a specified density, total column, metallicity, and relative abundances of carbon and nitrogen to oxygen, the model can then produce predicted column densities for \ion{H}{1} and metal ions, $N_{\rm X}(n_{\rm H}, N_{\rm H}, {\rm [O/H], [C/O], [N/O]})$. We fit the observed column densities for the main component and wing with the predicted PIE models using \texttt{LMFIT} and explored the posterior with \texttt{emcee}. 
The posteriors for the ionization model fits to the main component and wing are summarized in a corner plot and violin plot in Figure \ref{figure:cloudy}.

The main component's \ion{H}{1}, oxygen \textcolor{black}{ion, and sulfur ion} column densities can be adequately described by a single phase photoionization equilibrium model \textcolor{black}{with solar [S/O]} with a density of $\log n_{\rm H}/{\rm cm}^{-3}=-3.7 \pm 0.1$, total column of $\log N({\rm H})/{\rm cm}^{-2}=18.6 \pm 0.1$, metallicity of $\rm [O/H]=-0.6 \pm 0.1$, where the uncertainties are statistical given the single-density PIE model assumptions. However, models with solar relative abundances significantly overpredict the observed \ion{C}{3} column density as well as the \ion{N}{4} upper limit, resulting in inferred relative abundances of $\rm [C/O]=-0.6 \pm 0.1$ and a 95\% upper limit of $\rm [N/O]<-1.0$. The photoionization models that fit the main component column densities predict equilibrium temperatures of $\log T_{\rm PIE}/{\rm K}\approx 4.3$. If the highly-ionized, \ion{O}{6}-bearing absorber contributes nonnegligibly to the \ion{H}{1} column attributed to the main component, then the inferred metallicity would be proportionally higher and the total column proportionally lower, while the density and relative abundance estimates remain consistent within uncertainties. The cloud length-scale along the sightline implied by the inferred density and total column is $\log l/{\rm pc} = 3.8 \pm 0.2$. The inferred properties of the wing are broadly similar to the main component, though with substantially higher uncertainty. For the wing, the inferred properties are $\log n_{\rm H}/{\rm cm}^{-3}=-3.4 \pm 0.2$, $\log N({\rm H})/{\rm cm}^{-2}=17.6 \pm 0.2$, \textcolor{black}{$\log l/{\rm pc} = 2.4 \pm 0.3$}, $\rm [O/H]=-0.3 \pm 0.3$, $\rm [C/O]<-0.6$ and $\rm [N/O]<-0.3$. The volume-filling factor of the intermediate ionization-state gas along the quasar sightline is $f_{\rm V}\sim {l}/{L}\sim6\%$, where $l\approx 6$ kpc is the total length scale of the main component and wing and $L\approx 100$ kpc is the pathlength of the sightline through the halo. Assuming the intermediate-ion-bearing gas in this system is typical, the volume-filling factor averaged globally over dwarf CGM is $f_{\rm V}\sim \kappa(d<R_{\rm h})\times l/ \langle L \rangle \sim 2\%$, where $\kappa(d<R_{\rm h})\approx 0.3$ is the intermediate ion covering fraction at $d<R_{\rm h}$ from \cite{Mishra:2024} and $\langle L \rangle$ is the mean pathlength through the halo averaged over sightlines at $d<R_{\rm h}$.

To gain insights into the range of physical conditions that could produce the highly ionized \ion{O}{6} bearing gas with upper limits on associated \ion{O}{4}, \ion{S}{6}, and \ion{Ne}{8}, we ran a series of \texttt{Cloudy} models for gas with density ranging from $\log n_{\rm H}/{\rm cm^{-3}}=-4$ to $-6$ and temperature ranging from $\log T/{\rm K}=4$ to 6 for gas in steady-state between recombination and ionization from both photoionization by the UV background and collisional ionization. For models with $T=T_{\rm PIE}$, these correspond to photoionization equilibrium predictions. Models with $T>T_{\rm PIE}$ implicitly posit additional heating sources such as gravity or feedback, and they approach predictions for pure collisional ionization equilibrium at densities $n_{\rm H} \gg 10^{-4}\, {\rm cm}^{-3}$, where photoionization becomes subdominant.

Under collisional ionization equilibrium, the upper limit on the \ion{Ne}{8}/\ion{O}{6} and the lower limit on the  \ion{O}{6}/\ion{O}{4} combine to constrain the \ion{O}{6}-bearing CGM to have $\log T/{\rm K} \approx 5.4$ and $5.7$, $>0.4$ dex higher than the dwarf's estimated virial temperature. However, recent works have suggested that \ion{O}{6} absorption in dwarf CGM has significant contributions from gas photoionized by the UV background due to the low virial temperatures of low-mass halos \citep[e.g.,][]{{Johnson:2017, Qu:2018, Qu:2024, Dutta:2025a}}. To produce a strong \ion{O}{6} absorber with the measured limits on associated \ion{O}{4} and \ion{Ne}{8} from gas photoionized by the UV background, the implied gas density must be between $\log n_{\rm H}/{\rm cm}^{-3}\approx -5$ and $-4.4$. \textcolor{black}{For photoionized, O\,VI-bearing gas with a density of $\log n_{\rm H}/{\rm cm}^{-3}\lesssim -4.4$ and metallicity similar to the main component, the predicted H\,I column density is $N({\rm H\,I})<10^{14.2}\,\left[\frac{Z}{0.25 {\rm Z_\odot}}\right]^{-1}\,{\rm cm}^{-2}$. Consequently, uncertainty in the level of H\,I absorption associated with the highly ionized O\,VI absorption that may be misattributed to the main component introduces $\approx 0.1$ dex level systematic uncertainty in the main component H\,I column as well as inferred total column and metallicity. This systematic uncertainty is comparable to the statistical uncertainties estimated from the \texttt{Cloudy} and MCMC modeling.}

The predicted ion fraction of O$^{5+}$ ions in the photo$+$collisional ionization models do not exceed $f({\rm O}^{5+})=0.27$. We therefore adopt a fiducial ion fraction assumption of  $f({\rm O}^{5+})=0.2$ for consistency with past work \citep[e.g.,][]{Tumlinson:2011} to estimate the total oxygen column associated with the highly ionized component of $N({\rm O}) \sim 10^{15}\ {\rm cm^{-2}}$. Assuming that the highly ionized CGM exhibits a metallicity similar to the less ionized gas observed in the main component of $[\rm O/H]\approx-0.6$, this corresponds to a total hydrogen column density of

\begin{displaymath}
N({\rm H}) \sim 10^{19}\,\left[ \frac{f({\rm O^{5+}})}{0.2} \right]^{-1} \left[ \frac{Z}{0.25\, {\rm Z_\odot}} \right]^{-1}\ {\rm cm^{-2}}
\end{displaymath}
and implied length-scale of
\begin{displaymath}
l \sim 100 \left[ \frac{f({\rm O^{5+}})}{0.2} \right]^{-1} \left[ \frac{Z}{0.25\, Z_\odot} \right]^{-1} \left[ \frac{n_{\rm H}}{2\times 10^{-5}\,{\rm cm^{-3}}} \right]^{-1}\ {\rm kpc,}
\end{displaymath} where the density normalization corresponds to the density at which the $\rm O^{5+}$ ion fraction peaks under photoionization equilibrium. We note that this density is subject to factor of $\sim 2$ uncertainty from potential local contributions by soft X-ray sources that could contribute to the UV background along the quasar sightline \citep[][]{Werk:2016, Upton-Sanderbeck:2018} and even reduce cooling in low-mass halos \citep[][]{Cantalupo:2010}. Even so, the inferred length scale is comparable to the path length of the sightline through the dwarf galaxy's halo, suggesting that the \ion{O}{6} absorption may arise from an approximately volume-filling photoionized medium. Finally, the \ion{Ne}{8} nondetection indicates that any hot, $T\approx 10^6$ K phase of the CGM has a total metal column density less than \ion{O}{6}-bearing gas.

\section{Discussion}
\label{section:discussion}
\subsection{Comparison with previous dwarf CGM surveys}
Recent surveys of the CGM of isolated, star-forming dwarf galaxies at intermediate redshifts \citep[][]{Mishra:2024} found covering fractions for strong \ion{H}{1} and \ion{O}{6} systems of $\log N/{\rm cm}^{-2} > 14$ at projected distances of  $d<0.5 R_{\rm h}$ of $\approx 100\%$ and $\approx 80\%$, respectively, suggesting that the \ion{H}{1} and \ion{O}{6} absorption levels detected around the dwarf galaxy in this work are fairly common. At low redshifts of $z\lesssim 0.05$, the \ion{C}{3}, \ion{O}{3}, and \ion{O}{4} features observed here are not accessible. To compare the CGM properties of the dwarf in the field of \HE\ with surveys in the local Universe, we therefore used the ionization models for the main component and wing described in Section \ref{section:ionization} to make predictions for the \ion{Si}{3}, \ion{Si}{4}, and \ion{C}{4} columns which are observable at low redshift. Assuming $\rm [Si/O]=0$ and the measured carbon-to-oxygen for the main component, the total \ion{Si}{3}, \ion{Si}{4}, and \ion{C}{4} columns expected in the dwarf galaxy's CGM are $\log N/{\rm cm}^{-2}\approx12.5$, $12.6$, and $13.2$, respectively. While the column density expected in \ion{Si}{3} is high enough to be detected in the majority of low-$z$ dwarf galaxy CGM sightlines compiled by \cite{Zheng:2024}, the expected \ion{Si}{4} and \ion{C}{4} columns are comparable to typical detection limits. When the low$-z$ surveys are restricted to sightlines with sufficient $S/N$ to detect absorption at these levels, the estimated covering fractions for such \ion{Si}{3}, \ion{Si}{4}, and \ion{C}{4} absorbers are $\approx50\%$ at $d\lesssim 0.5 R_{\rm h}$, after converting to a consistent stellar-to-halo mass relation and virial radius definition. The CGM properties observed around the dwarf galaxy in the field of \HE\ are therefore fairly common given its stellar mass and the projected distance from the quasar sightline.

\subsection{Physical conditions of the dwarf CGM}
The physical picture needed to describe the combination of observed intermediate and high ions in the CGM of dwarfs is not well established. Dwarf CGM could consist of a warm-hot, volume-filling medium populated by cool, photoionized clouds in approximate pressure equilibrium, similar to massive galaxies \citep[e.g.,][]{Zahedy:2019}. However, while Jean's stable, such warm-hot CGM is expected to cool rapidly in low-mass halos, so that turbulence or bulk motion may dominate over thermal pressure \citep[e.g.,][]{Fielding:2017, Correa:2018, Stern:2020, Lochhaas:2020, Gurvich:2023, Kakoly:2025, Cook:2025}. In this case, the intermediate ions and highly ionized \ion{O}{6} could arise from cool, photoionized gas of different densities tracing an approximately lognormal density distribution induced by turbulence. Moreover, gas consistent with each of these two pictures may be present in different parts of the same halo.
\textcolor{black}{The temperature inferred for the main component from Voigt profile modeling of $\log T/{\rm K}=4.6 \pm 0.2$ is somewhat higher than the photoionization equilibrium temperature of $\log T_{\rm PIE}/{\rm K}\approx4.3$ expected from the ionization modeling described Section \ref{section:ionization}, though consistent within uncertainty at the 1.5$\sigma$ level. In principle, heating from turbulent dissipation or feedback could result in $T>T_{\rm PIE}$. However, the two temperatures are consistent statistically, and the temperature estimated from Voigt profile fitting could be biased high by nonnegligible \ion{H}{1} contributions from the O\,VI-bearing CGM or additional, blended \ion{H}{1} components without detectable metal absorption. On the other hand, the temperature for the wing inferred from the Voigt profile models is within $0.1$ dex of the predicted PIE temperature.}

Around dwarfs, a volume-filling warm-hot medium could be virialized gas or a hot wind.  However, the star-formation rate surface density of the dwarf studied in this work is significantly lower than those of galaxies observed with fast, starburst outflows \citep[e.g.,][]{Heckman:2015}. Moreover, the \ion{Ne}{8} nondetection disfavors a significant hot phase. We therefore consider whether the \ion{O}{6} could arise from a warm, virialized phase with temperature $T_{\rm vir}\approx 10^5$ K. If in approximate pressure equilibrium with the cool, photoionized gas observed in the main component, a warm \ion{O}{6}-bearing phase would exhibit a density of $n_{\rm warm} \approx n_{\rm cool}T_{\rm PIE}/T_{\rm vir} \approx 4 \times 10^{-5}\ \rm cm^{-3}$. This density is similar to the peak density for photoionized \ion{O}{6} adopted in Section \ref{section:absorption}, indicating that photoionization likely plays a significant role, even if the \ion{O}{6} arises from warm, virialized gas.

If the \ion{O}{6} arises from a warm medium supported by a thermal pressure gradient, then thermal line-broadening is expected to dominate, leading to an expected \ion{O}{6} Doppler parameter of $b=10-15$\ \kms, significantly smaller than observed. The Doppler-width of the \ion{O}{6} bearing CGM of $b= 31$ \kms\ is, therefore, dominated by non-thermal broadening, and it is comparable to the estimated virial velocity of the dwarf, $v_{\rm vir}\approx 44$ \kms, which corresponds to a Doppler width of $b\approx v_{\rm vir}\sqrt{2/3}=35$ \kms. Without a clear detection of H\,I associated with the O\,VI-bearing CGM, the data cannot directly rule out a warm phase. However, if a significant fraction of the CGM exhibited both a temperature of $T\sim T_{\rm vir}$ and non-thermal broadening comparable to the virial velocity, then this would imply that the CGM is out of virial equilibrium and likely unbound, which is disfavored by the modest star-formation rate of the dwarf. This suggests that turbulent pressure or bulk motion are more important than thermal pressure in the highly ionized dwarf CGM. In this case, both the intermediate ions and strong \ion{O}{6} would arise from cool, photoionized gas, and they need not be in thermal pressure equilibrium.

\subsection{Mass budget}

The total mass in CGM probed by intermediate ions such as \ion{O}{4} can be estimated as $M\sim N({\rm H}) \pi R_h^2 \kappa(d<R_{\rm h})\mu m_{\rm p}$
where $\kappa$ is the covering fraction for similar absorption systems observed around dwarf galaxies of similar masses, $N({\rm H})$ is the total hydrogen column inferred for the main component, $\mu$ is the mean molecular weight, and $m_{\rm p}$ is the proton mass. Adopting the total column for the main component, $N({\rm H})\approx 4\times10^{18}\ {\rm cm^{-2}}$, and $\kappa(d<R_{\rm h}) \approx 0.3$ based on \cite{Mishra:2024} results in a mass estimate of
\begin{displaymath}
M\sim 2\times10^8 \left( \frac{N({\rm H})}{4\times10^{18}\ {\rm cm^{-2}}} \right) \left( \frac{\kappa}{0.3} \right) \left( \frac{R_{\rm h}}{60\ \rm kpc}\right)^2\ \rm M_\odot,
\end{displaymath}
representing $\approx 5\%$ of the dwarf galaxy's baryon budget.
Overall, this intermediate ionization state CGM mass estimate is fairly consistent with recent upper limits by \cite{Faerman:2025} based on \ion{H}{1} observations and estimates by \cite{Zheng:2024} based on total  low and intermediate ionization state metal column densities. The intermediate ionization state mass estimate is a factor of ten lower than estimates of the mass in the highly ionized \ion{O}{6}-bearing phase \citep{Mishra:2024, Dutta:2025a}. Despite being a small component of the baryon budget of the dwarf galaxy's halo, the intermediate ionization state CGM exceeds the stellar mass of the galaxy by a factor of four and is comparable to the expected ISM \ion{H}{1} mass \citep[e.g.,][]{Parkash:2018}.

\begin{figure*}
\includegraphics[width=\textwidth]{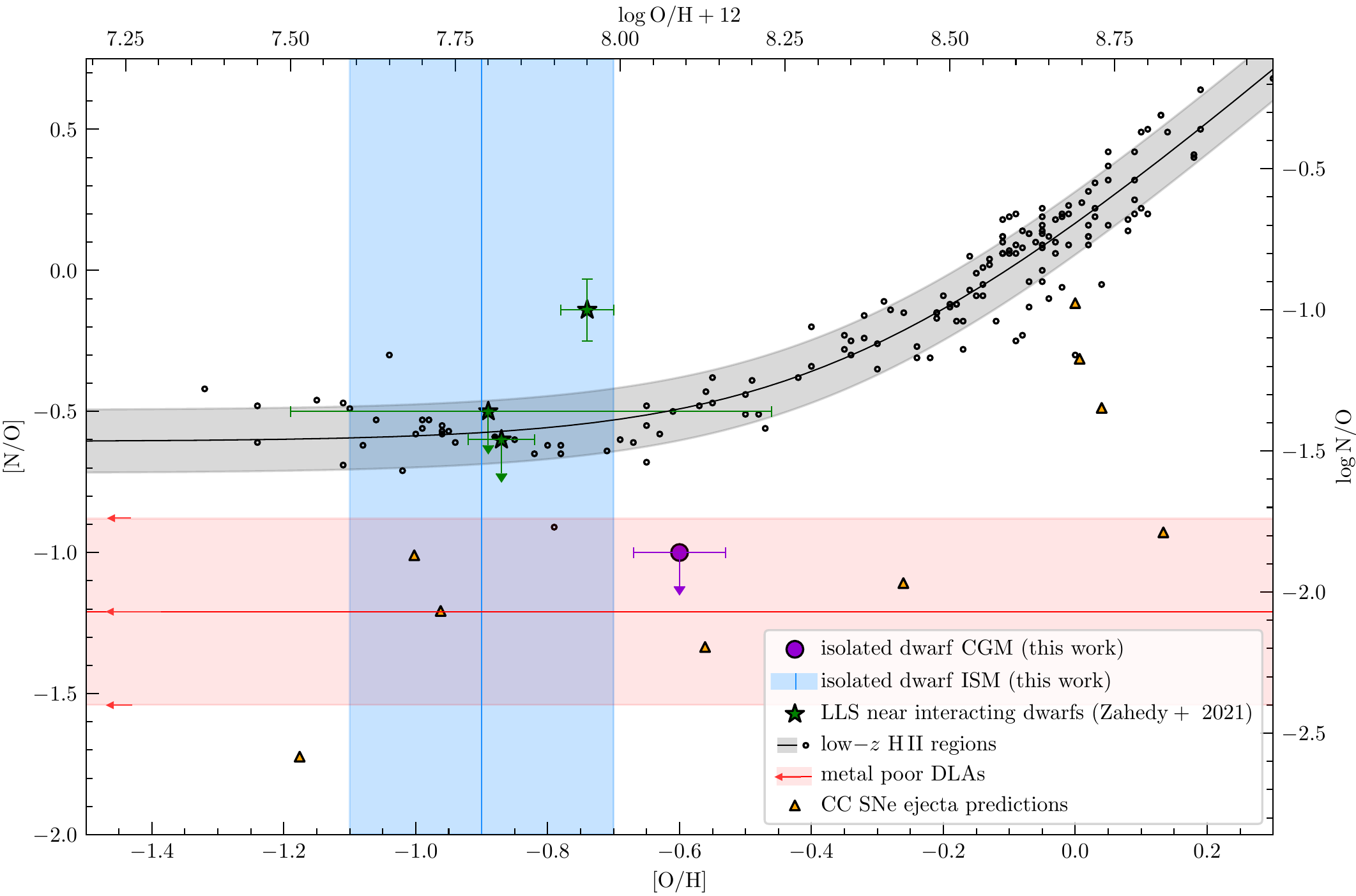}
\caption{Relative abundance of nitrogen to oxygen versus oxygen metallicity for the CGM of the isolated dwarf galaxy along the sightline to \HE\ at $z=0.5723$ (dark violet circle) compared to the ISM of galaxies measured in \ion{H}{2} regions (black circles with a black line showing the mean curve) from \cite{Esteban:2009, Berg:2012, Pilyugin:2014, Berg:2016}. The figure also shows the metallicity and [N/O] abundances (green stars) from \citealt{Zahedy:2021} for components of a Lyman Limit System (LLS) found near an interacting system of dwarfs \citep[][]{Chen:2020}. The $1\sigma$ scatter about the mean curve relating [N/O] abundance to oxygen metallicity for the \ion{H}{2} regions is shown as a grey band. The estimated oxygen metallicity of the ISM of the dwarf galaxy in the field of \HE\ is marked by a vertical blue line with a faded band representing measurement error. The metallicity of the dwarf galaxy's CGM is comparable to that of its ISM, but with an [N/O] abundance upper limit that is \textcolor{black}{0.4} dex below expectations for the ISM of galaxies with similar mass and metallicity. On the other hand, the low nitrogen abundance of $\rm [N/O]<-1.0$ is consistent with those of metal-poor DLAs calculated by \cite{Nunez:2022}, who interpreted the ratio as representing the abundances of ejecta from low-metallicity core-collapse supernovae. The mean [N/O] ratio of the low metallicity DLAs is marked by a red line with a faded band marking $1\sigma$ scatter, though we note that the metallicities of these DLAs are below the plotted range, as indicated with leftward arrows. For comparison, the predicted [N/O] abundances of IMF averaged ejecta from core-collapse supernovae from \cite{Johnson:2023} are shown in orange triangles versus progenitor metallicity. The relatively high metallicity and low [N/O] ratio of the dwarf galaxy CGM are consistent with the gas originating from outflows driven by core-collapse supernovae.}
\label{figure:abundances}
\end{figure*}

\subsection{Chemical abundances and dwarf CGM origins}
\label{section:abundances}
Oxygen is primarily produced by massive stars and ejected into the ISM when they explode as core-collapse supernovae shortly after star formation. On the other hand, a significant fraction of carbon and the overwhelming majority of nitrogen in the ISM arise from mass loss by evolved, low-mass stars in their giant phases, $\sim 10^8$ years after star formation \citep[e.g.,][]{Kobayashi:2020}. Consequently, the abundances of carbon and nitrogen relative to oxygen can serve as a diagnostic of the nucleosynthetic history of material in and around galaxies \citep[][]{Tinsley:1979}. For example, ionized absorption systems in the CGM/IGM at $z<1$ sometimes exhibit super-solar [N/O] and [C/O], consistent with chemically evolved gas expected in the ISM of massive galaxies where oxygen rich core-collapse supernovae ejecta and carbon/nitrogen rich ejecta from giant stars are effectively mixed  \citep[][]{Zahedy:2021, Kumar:2024}, but with metallicities lower than is observed in the ISM of massive galaxies. Such chemically evolved but modest-metallicity gas can be readily explained if it is ejected from the ISM of Milky Way-like galaxies into the CGM and then mixes with more chemically pristine IGM/CGM to reduce overall metallicity without significantly changing the relative [C/O] and [N/O] abundances.

In contrast, the CGM of the dwarf galaxy in the field of \HE\ exhibits low relative abundances  of $\rm [C/O]=-0.6$ and $\rm [N/O]<-1.0$ despite a relatively high metallicity of $\rm [O/H]=-0.6$, comparable to the metallicity estimated for the dwarf's ISM. To interpret the likely origins of gas with this chemical signature, we collected measurements of [N/O] and [O/H] of \ion{H}{2} regions of both massive galaxies and low-mass dwarfs \citep[][]{Esteban:2009, Berg:2012, Pilyugin:2014, Berg:2016} and compare them to the [N/O] abundance for the dwarf galaxy's CGM in Figure \ref{figure:abundances}. The \ion{H}{2} regions for dwarf galaxies generally exhibit relatively low metallicities of ${\rm [O/H]}<-0.5$ and sub-solar nitrogen-to-oxygen ratios of $\rm [N/O]\approx -0.6$ with an intrinsic scatter of $\approx 0.1$ dex. This expected ISM [N/O] abundance is $0.4$ dex above the [N/O] upper limit inferred for the dwarf's CGM. The $\rm [N/O]<-1.0$ limit for the dwarf's CGM is consistent with the relative abundances of low-metallicity damped Ly$\alpha$ absorbers \citep[DLAs; e.g.,][]{Petitjean:2008, Pettini:2008}, which exhibit a mean ratio of $\rm [N/O]=-1.2$ with $0.3$ dex scatter \citep[][]{Nunez:2022}. The low [N/O] relative abundances of low-metallicity DLAs are often interpreted as being due to core-collapse supernovae ejecta relative abundances in gas that has not been significantly enriched in nitrogen from evolved, low-mass stars. Indeed, when integrated over the IMF, simulations of low-metallicity core-collapse supernovae \citep[][]{Woosley:1995, Nomoto:2013, Limongi:2018} suggest that their ejecta will exhibit abundance ratios in the range of $\rm -1.7<[N/O] < -0.9$ \citep[][]{Johnson:2023}. \textcolor{black}{Similarly, the carbon abundance of the main component, $\rm [C/O]=-0.6$, is also below the mean expected from the ISM of dwarf galaxies of similar mass and metallicity of ${\rm [C/O]} \approx -0.4$ \citep[e.g.,][]{Garnett:1995, Berg:2016}. However, we caution that measured [C/O] abundance ratios of H\,II regions in dwarf galaxies exhibit higher dispersion of $\approx 0.25$ dex and are subject to different systematics for measurements using collisionally excited lines versus recombination lines \citep[see discussion in][]{Berg:2016}.}

The low nitrogen-to-oxygen ratio and relatively high oxygen metallicity of the CGM of the dwarf galaxy can therefore be explained if the gas arises from outflows driven by core-collapse supernovae, but without significant mixing with more nitrogen enriched ISM. Such limited mixing of a supernova outflow with the ISM suggests a low mass-loading factor, raising the possibility that feedback may be primarily energy mediated \citep[e.g.,][]{Carr:2023, Voit:2024}, though additional exploration of these models in the low-mass regime is needed. In this case, the oxygen rich core-collapse supernovae ejecta reaches the CGM without mixing significantly with more nitrogen rich ISM, and the moderate metallicity of the absorber can be explained by subsequent mixing with more pristine CGM/IGM. \textcolor{black}{Core-collapse supernovae winds escaping to the CGM after only limited mixing with ISM may produce an anti-correlation between ISM [N/O] abundance and wind mass-loading \citep[e.g.,][]{Arellano-Cordova:2025}}. Alternatively, the oxygen enriched gas could have been ejected at significantly earlier times, before the ISM was enriched in nitrogen.

To our knowledge, the only other similarly low-mass halo with constraints on metallicity and relative abundances of ionized CGM arises near an interacting system of dwarfs discovered around a Lyman Limit System \citep[LLS;][]{Chen:2020}. While the metallicity of the components in this LLS are similar to the dwarf CGM studied here, \citealt{Zahedy:2021} found higher [N/O] abundance in one component, which may arise from more chemically mature ISM being stripped into the CGM during the dwarf$-$dwarf interaction. Together, these systems showcase the diverse origins of CGM in low-mass halos, which can be constrained with measurements of metallicity and relative abundances.


\section{Summary and Conclusions}
\label{section:conclusions}

In this Letter, we presented the first study of the physical conditions and multi-element abundances in ionized, absorbing CGM around a low-mass, isolated, star-forming dwarf galaxy at intermediate redshift ($z=0.5723$), with a stellar mass of $M_\star \approx 5\times10^7\ {\rm M_\odot}$ in the field of a bright background quasar \HE, with archival UV spectra from {\it Hubble}. The dwarf galaxy exhibits inferred star-formation rates and ISM metallicity consistent with expectations for galaxies of similar mass. Deep and highly complete galaxy redshift surveys in the field demonstrate that the dwarf resides in an isolated environment, more than 2 Mpc from the nearest luminous galaxy.

The CGM of the dwarf is detected in absorption from \ion{H}{1} and multiple intermediate metal ions (\ion{O}{3}, \ion{O}{4}, \ion{C}{3}) as well as kinematically broader, highly ionized \ion{O}{6}. Further, the data enable stringent upper limits on nitrogen ions and \ion{Ne}{8}. Photoionization analysis of the intermediate ionization state gas indicates that it is consistent with arising in cool (${\sim}10^4$ K) CGM with a modest volume-filling factor of $\sim 6\%$ along the sightline and a total mass that is comparable to the ISM \ion{H}{1} mass of similar dwarf galaxies, but substantially less than the mass in the highly ionized CGM phase detected in strong \ion{O}{6}. The virial temperature of the dwarf galaxy is below levels required to produce significant \ion{O}{6} absorption via collisional ionization, suggesting that photoionization by the UV background plays a significant role. In this scenario, the highly ionized CGM is approximately volume filling and may be warm, virialized gas in approximate gas pressure equilibrium with the gas detected in intermediate ions. However, the kinematics of the \ion{O}{6} absorption are comparable to the virial velocity, suggesting that bulk motion or turbulence dominate over thermal pressure. This suggests the intermediate ionization state absorption and highly ionized \ion{O}{6} could both arise from cool, photoionized gas tracing different parts of a broad, lognormal density distribution predicted in turbulent CGM \citep[e.g.,][]{Stern:2021, Kakoly:2025}.

The \ion{H}{1} and intermediate ion measurements enable new insights into the chemical composition of the dwarf's CGM, which exhibits an inferred oxygen metallicity of $\rm [O/H]\approx-0.6$, comparable to the metallicity estimated for ISM of the dwarf galaxy. However, the inferred relative nitrogen-to-oxygen abundance of $\rm [N/O]<-1.0$ is well below levels expected from the dwarf's current ISM but consistent with expectations from core-collapse supernovae ejecta. This suggests that the oxygen enriched CGM escaped from the dwarf without significant mixing with ISM enriched in nitrogen from evolved, low-mass stars.

\section*{Acknowledgments}
\textcolor{black}{This Letter is dedicated to the memory and legacy of Fakhri S. Zahedy. It was inspired by his pioneering work on relative abundances in the CGM of massive galaxies and Lyman limit systems and would not have been possible without his creative vision and scientific insights.}

We thank Joop Schaye, Marijke Segers, Lorrie Straka, and Monica Turner for their contributions to and leadership of the MUSEQuBES project. Based on observations from the European Organization for Astronomical Research in the Southern Hemisphere under ESO, the 6.5 meter Magellan Telescopes located at Las Campanas Observatory, Chile, and the NASA/ESA Hubble Space Telescope operated by the Space Telescope Science Institute (STScI) under the Association of Universities for Research in Astronomy, Inc. NASA contract (NAS 5–26555).
This Letter made use of the NASA/IPAC Extragalactic Database and the NASA Astrophysics Data System. 
SDJ, NM, EF, ZL, and AK acknowledge partial support from STScI grants HST-GO-15935.012-A, HST-GO-16728.001-A, HST-GO-17517.013-A, HST-GO-17800.001-A, and HST-GO-17803.001-A.
CAFG was supported by NSF through grants AST-2108230 and AST-2307327; by NASA through grants 21-ATP21-0036 and 23-ATP23-0008; and by STScI through grant JWST-AR-03252.001-A.
JS is supported by a grant from the US-Israel Binational Science Foundation.
SC gratefully acknowledges support from the European Research Council (ERC) under the European Union’s Horizon 2020 Research and Innovation programme grant agreement No 864361.
ZQL is supported by NASA under the Future Investigators in NASA Earth and Space Science and Technology (FINESST) program through grant number 24-ASTRO24-0070.

\section*{Software}
\texttt{Astropy} \citep[][]{Astropy-Collaboration:2022},
\texttt{Bagpipes} \citep[][]{Carnall:2018},
\texttt{Cloudy}  \citep[][]{Chatzikos:2023},
\texttt{emcee} \citep[][]{Foreman-Mackey:2013},
 \texttt{LINMIX} \citep[][]{Kelly:2007},
\texttt{LMFIT}  \citep[][]{Newville:2016},
\texttt{MATPLOTLIB} \cite[][]{Hunter:2007},
\texttt{MPDAF} \cite[][]{Bacon:2016, Piqueras:2017},
\texttt{Numpy} \citep[][]{Harris:2020},
\texttt{Scipy} \citep[][]{Virtanen:2020}.
\section*{Data Availability}
The MUSE data are available in the ESO archive (PI: Schaye, PID: 094.A-0131). The {\it HST} COS spectra and ACS images are available in MAST (\dataset[10.17909/b4m3-yd95]{http://dx.doi.org/10.17909/b4m3-yd95}).
\facilities{HST (ACS, COS), VLT:Yepun (MUSE), Magellan:Clay (LDSS3), Magellan:Baade (IMACS), and Blanco (DECam)}

\appendix
\section{R23$-$metallicity calibration}
\label{appendix:R23}
\begin{figure*}
\includegraphics[width=\textwidth]{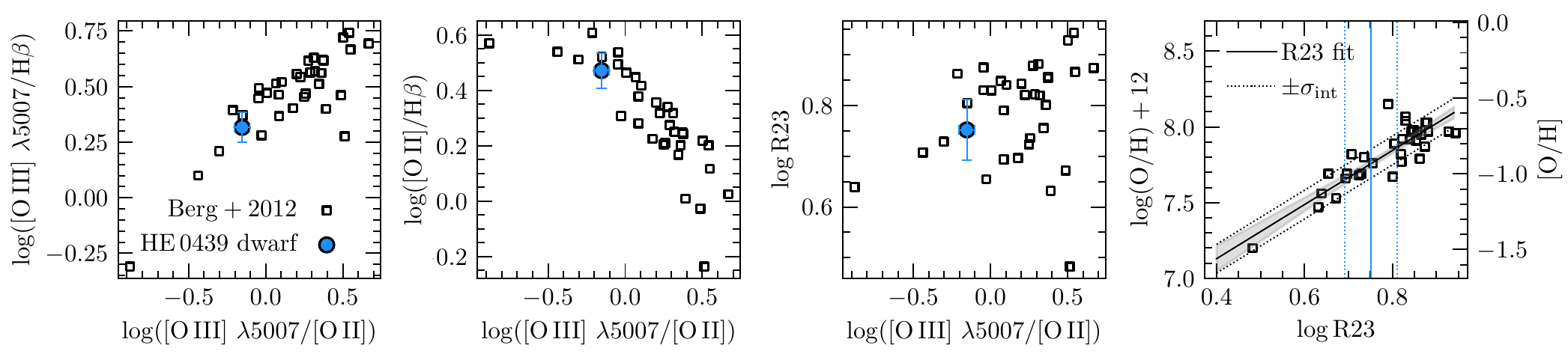}
\caption{Emission line ratios of the dwarf galaxy at $z=0.5723$ in the field of \HE\ (blue circle) compared to \ion{H}{2} regions in local dwarf galaxies of $\log M_\star/{\rm M_\odot} <9$ observed by \cite{Berg:2012} (black squares). The {\it right} panel displays \ion{H}{2} region metallicities for the local dwarf galaxies versus the metallicity-sensitive R23 diagnostic line ratio. The measured R23 ratio  for the dwarf in the field of \HE\ is marked by a vertical, solid blue line with dashed lines marking the $1\sigma$ uncertainty. A best-fit power-law relationship between R23 and metallicity, described in Appendix \ref{appendix:R23}, is shown in black solid line with 68\% credible interval and intrinsic scatter shown in a grey band and dotted line, respectively.}
\label{figure:R23}
\end{figure*}

\restartappendixnumbering
The R23 ratio is often used to estimate extragalactic \ion{H}{2} region metallicities \citep[for a review, see][]{Maiolino:2019}, particularly for low-mass galaxies where priors from the mass-metallicity relation mitigates uncertainty associated with the double-valued nature of the R23$-$metallicity relation. However, recent R23 diagnostic calibrations have focused on galaxies with high-ionization states to enable use at high redshift \citep[e.g.,][]{Jiang:2019}, which may not be appropriate for systems with modest ionization states like the dwarf galaxy studied here. Consequently, we recalibrated the R23$-$metallicity diagnostic for dwarf galaxies with a range of ionization states. First, we identified \ion{H}{2} regions in local dwarf galaxies of $\log M_\star/{\rm M_\odot<9}$ with direct $T_e$-based metallicities from \cite{Berg:2012}. Figure \ref{figure:R23} displays the strong line ratios of these \ion{H}{2} regions along with those measured for the dwarf galaxy in this work, demonstrating that the local sample includes systems with similar line ratios. We then plotted the measured oxygen metallicity for the dwarf \ion{H}{2} regions from \cite{Berg:2012} versus the measured $\log \rm R23$ line ratio. The measured metallicities are well correlated with R23, and we fit a linear model to this relation, $\log {\rm O/H}+12 = a + b \log {\rm R23}$ with intrinsic scatter $\sigma_{\rm int}$ using the \texttt{LINMIX} package \citep[][]{Kelly:2007}, resulting in $a=6.4 \pm 0.2, b=1.8 \pm 0.2,$ and $\sigma_{\rm int}=0.1 \pm 0.02$. This relation is only valid for dwarf galaxies on the low-metallicity branch of the metallicity$-$R23 relation.

\section{Contaminating absorption modeling}
\label{appendix:contamination}

\subsection{Contamination of the dwarf CGM \ion{H}{1} Ly$\beta$ by \ion{H}{1} Ly$\alpha$ at z=0.328}

\begin{figure*}
\includegraphics[width=\textwidth]{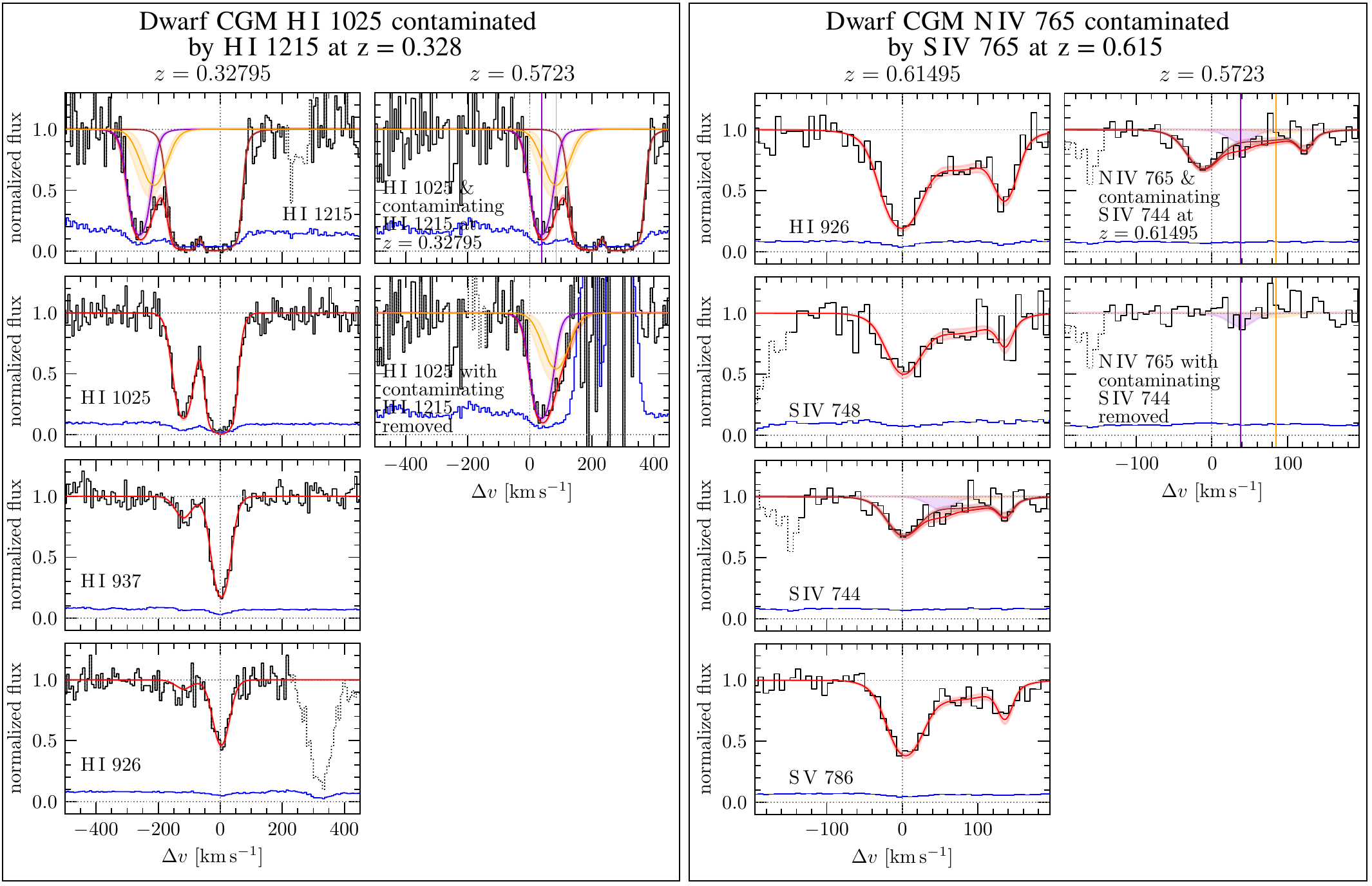}
\caption{\textcolor{black}{Continuum normalized spectra and models of absorption contaminating the dwarf CGM features. The {\it left} set of panels display  \ion{H}{1} Ly$\alpha$ at $z=0.328$ that contaminates the dwarf galaxy CGM \ion{H}{1} Ly$\beta$ absorption. The {\it right} set of panels display  \ion{S}{4} $\lambda$744 at $z=0.615$ that contaminates the dwarf galaxy CGM \ion{N}{4} $\lambda765$ spectral region. In each set, the four panel column on the {\it left} displays the contaminating system in the top panel with absorption features that constrain it in the lower three panels with velocity zeropoint set by the redshift of the strongest component of contaminating system, which is labeled at the top of the column. The two panel columns on the {\it right} display the dwarf CGM features and contamination in the {\it top} panel and the dwarf CGM absorption after dividing out by the contamination model in the panel below, with the velocity zeropoint set by the dwarf galaxy redshift, which is also labeled at the top of the column. In each panel, the full model including both the dwarf CGM and contaminating features is shown in red, contamination model is shown in brown, and absorption for the dwarf CGM main component and wing are shown in dark violet and orange. For each model, the dark line indicates the best-fit while the faded band marks the 68\% credibility interval. The details of the modeling of the contamination is described in Appendix \ref{appendix:contamination}.}}
\label{figure:contamination}
\end{figure*}

\textcolor{black}{The \ion{H}{1} Ly$\beta$ absorption associated with the CGM of the dwarf is contaminated by \ion{H}{1} Ly$\alpha$ from a strong H\,I absorption system at $z=0.328$, particularly impacting the red side of the wing component. While several of the Lyman series lines of the $z=0.328$ system are themselves contaminated by other absorbers, the strength of the \ion{H}{1} Ly$\alpha$ line is well constrained by absorption from Ly$\beta$, Ly$\epsilon$, and Ly$\eta$. The strong H\,I system at $z=0.328$ can be adequately modeled by two components with shared redshift and Doppler width. To infer the H\,I absorption properties associated with the dwarf CGM while accounting for increased uncertainty associated with the contamination, we fit a two-component Voigt profile model to the \ion{H}{1} at $z=0.328$ simultaneously with the $\chi^2$ minimization and MCMC fitting of the dwarf's CGM, as described in Section \ref{section:absorption}. The inferred properties of the weaker, bluer Voigt component fit to the contaminating H\,I are $z_1=0.32742 \pm 0.00001$, $b_1=32\pm2$, and $\log N_1({\rm H\,I})/{\rm cm}^{-2}=14.82\pm0.02$. The inferred properties of the stronger, redder Voigt component fit to the contaminating \ion{H}{1} are 
$z_2=0.32795 \pm 0.00001$, $b_2=29\pm2$, and $\log N_2({\rm H\,I})/{\rm cm}^{-2}=15.81\pm0.02$. The results of the simultaneous fitting to the dwarf CGM and contaminated H\,I as well as the dwarf CGM after dividing by the best-fitting contamination model are shown in Figure \ref{figure:contamination}.}

\subsection{Contamination of the dwarf CGM \ion{N}{4} by \ion{S}{4} $\lambda$744 at z=0.615}

\textcolor{black}{The expected observed wavelength of \ion{N}{4} absorption associated with the CGM of the dwarf is contaminated by \ion{S}{4} $\lambda$744 from an absorption system at $z=0.615$ that is detected in \ion{H}{1} and a wide array of heavy element ions. The contamination level is well constrained by detection of \ion{S}{4} $\lambda$748, which has an oscillator strength two times greater than the $\lambda$744 transition. To infer the level of N\,IV in the dwarf CGM allowed by the data, we fit the contaminating \ion{S}{4}, with component structure also guided by \ion{H}{1} and \ion{S}{5} associated with the $z=0.615$ system. The system exhibits two distinct components separated by $\approx 70$ \kms, but with nonnegligible absorption between them that can be modeled with an additional, broader component. We therefore fit the \ion{H}{1}, \ion{S}{4}, and \ion{S}{5} features for the $z=0.615$ system with a three component model, with components sharing common redshift, non-thermal broadening, and temperature, while also allowing for \ion{N}{4} $\lambda765$ absorption associated with the main component and wing in the dwarf CGM. We performed $\chi^2$ minimization and MCMC exploration of the posterior as described in Section \ref{section:absorption}, and simultaneously varied model parameters associated with both absorption systems. The inferred properties for the strongest, bluer component are $z_1=0.61495\pm0.00001$, $b_{\rm nt,1}=24\pm2$ \kms, and $\log T_{1}/{\rm K}=4.0 \pm 0.14$, with column densities of $\log N_{\rm 1}({\rm H\,I})/{\rm cm}^{-2}=16.13\pm0.04$,
$\log N_{\rm 1}({\rm S\,IV})/{\rm cm}^{-2}=13.61\pm0.05$, and $\log N_{\rm 1}({\rm S\,V})/{\rm cm}^{-2}=13.27\pm0.04$. The inferred properties for the weaker, redder component are $z_3=0.61568 \pm 0.00001$, $b_{\rm nt,3}=8\pm3$ \kms, and $\log T_{3}/{\rm K}=4.0 \pm 0.1$, with column densities of $\log N_{\rm 3}({\rm H\,I})/{\rm cm}^{-2}=15.66\pm0.07$,
$\log N_{\rm 3}({\rm S\,IV})/{\rm cm}^{-2}=13.0\pm0.15$, and $\log N_{\rm 3}({\rm S\,V})/{\rm cm}^{-2}=12.6\pm0.1$. While the temperature for the broader component is not well constrained, the inferred redshift and line-width are $z_2=0.6153 \pm 0.0001$ and $b_{2}=55\pm10$ \kms, with column densities of $\log N_{\rm 2}({\rm H\,I})/{\rm cm}^{-2}=15.68\pm0.08$,
$\log N_{\rm 2}({\rm S\,IV})/{\rm cm}^{-2}=13.31\pm0.12$, and $\log N_{\rm 1}({\rm S\,V})/{\rm cm}^{-2}=12.84\pm0.10$. The results of the simultaneous fitting to the dwarf CGM and contaminated S\,IV are shown in Figure \ref{figure:contamination}.}

\bibliography{manuscript.bib}
\bibliographystyle{aasjournalv7}



\end{document}